\RequirePackage{fix-cm}
\RequirePackage{fixltx2e}
\documentclass[english]{scrartcl}
\usepackage[T1]{fontenc}
\usepackage[latin9]{inputenc}
\usepackage{geometry}
\usepackage[active]{srcltx}
\usepackage{color}
\usepackage{babel}
\usepackage{array}
\usepackage{float}
\usepackage{booktabs}
\usepackage{textcomp}
\usepackage{enumitem}
\usepackage{multirow}
\usepackage{amsmath}
\usepackage{amsthm}
\usepackage{amssymb}
\usepackage{stackrel}
\usepackage{graphicx}
\usepackage[authoryear]{natbib}
\usepackage[unicode=true,pdfusetitle,
 bookmarks=true,bookmarksnumbered=true,bookmarksopen=true,bookmarksopenlevel=1,
 breaklinks=true,pdfborder={0 0 1},backref=false,colorlinks=true]
 {hyperref}
\hypersetup{
 linkcolor=blue, citecolor=blue, urlcolor=blue, filecolor=blue,pdfpagelayout=OneColumn, pdfnewwindow=true,pdfstartview=XYZ, plainpages=false, pdfpagelabels, hyperindex=true}
\usepackage{breakurl}

\makeatletter

\providecommand{\tabularnewline}{\\}

\theoremstyle{definition}
\newtheorem{condition}{\protect\conditionname}
\theoremstyle{plain}
\newtheorem{lem}{\protect\lemmaname}[section]

\usepackage{fancyhdr}
\usepackage{lastpage}
\usepackage{amsmath}
\usepackage{graphicx}
\usepackage{color}
\usepackage{fancybox} 
\usepackage{moreverb} 
\usepackage{listings} 
\usepackage{backref}
\usepackage {afterpage}

\usepackage{ifpdf} % part of the hyperref bundle
\ifpdf % if pdflatex is used

 \IfFileExists{lmodern.sty}{\usepackage{lmodern}}{}

\fi % end if pdflatex is used

\let\myTOC\tableofcontents
\renewcommand\tableofcontents{%
  \pdfbookmark[1]{\contentsname}{}
  \myTOC
}

\def\LyX{\texorpdfstring{%
  L\kern-.1667em\lower.25em\hbox{Y}\kern-.125emX\@}
  {LyX}}

\@addtoreset{footnote}{section}

\renewcommand*{\backref}[1]{}
\renewcommand*{\backrefalt}[4]{%
   \ifcase #1 %no citation
    \or %cited on exactly one page
      (Cited on page~#2)%
   \else%cited on multiple pages
      (Cited on pages~#2)
    \fi} 

\usepackage{dsfont}
\usepackage{bbm}

\usepackage[algo2e,ruled,vlined]{algorithm2e}
\usepackage{xcolor}
\definecolor{algoColorKeyword}{named}{blue}
\definecolor{algoColorComment}{named}{olive}

\SetKwComment{Comment}{$\triangleright$\ }{}

\makeatother

\providecommand{\conditionname}{Condition}
\providecommand{\lemmaname}{Lemma}

\begin{document}

\title{Fast and stable multivariate kernel density estimation by fast sum
updating}

\author{Nicolas Langren\'e\thanks{CSIRO Data61, RiskLab Australia, nicolas.langrene@csiro.au},
Xavier Warin\thanks{EDF R\&D, FiME (Laboratoire de Finance des March\'es de l'\'Energie),
warin@edf.fr}}

\date{\vspace{0.25em}
First version: December 5, 2017\\
\vspace{0em}
This version: October 22, 2018\\
\vspace{0.75em}
Accepted for publication in the\\
\vspace{0em}
\textit{Journal of Computational and Graphical Statistics}\vspace{0em}
}
\maketitle
\begin{abstract}
\noindent Kernel density estimation and kernel regression are powerful
but computationally expensive techniques: a direct evaluation of kernel
density estimates at $M$ evaluation points given $N$ input sample
points requires a quadratic $\mathcal{O}(MN)$ operations, which is
prohibitive for large scale problems. For this reason, approximate
methods such as binning with Fast Fourier Transform or the Fast Gauss
Transform have been proposed to speed up kernel density estimation.
Among these fast methods, the Fast Sum Updating approach is an attractive
alternative, as it is an exact method and its speed is independent
of the input sample and the bandwidth. Unfortunately, this method,
based on data sorting, has for the most part been limited to the univariate
case. In this paper, we revisit the fast sum updating approach and
extend it in several ways. Our main contribution is to extend it to
the general multivariate case for general input data and rectilinear
evaluation grid. Other contributions include its extension to a wider
class of kernels, including the triangular, cosine and Silverman kernels,
its combination with parsimonious additive multivariate kernels, and
its combination with a fast approximate k-nearest-neighbors bandwidth
for multivariate datasets. Our numerical tests of multivariate regression
and density estimation confirm the speed, accuracy and stability of
the method. We hope this paper will renew interest for the fast sum
updating approach and help solve large-scale practical density estimation
and regression problems.\vspace{2em}

\vspace{2em}


\newpage{}
\end{abstract}

\section{Introduction}

Let $(x_{1},y_{1}),(x_{2},y_{2}),\ldots,(x_{N},y_{N})$ be a sample
of $N$ input points $x_{i}$ and output points $y_{i}$ drawn from
a joint distribution $(X,Y)$. The kernel density estimator (aka Parzen-Rosenblatt
estimator) of the density of $X$ at the evaluation point $z$ is
given by:
\begin{equation}
\hat{f}_{\mathrm{KDE}}(z):=\frac{1}{N}\sum_{i=1}^{N}K_{h}(x_{i}-z)\label{eq:localdens}
\end{equation}

where $K_{h}(u):=\frac{1}{h}K\!\left(\frac{u}{h}\right)$ with kernel
$K$ and bandwidth $h$. The Nadaraya-Watson kernel regression estimator
of $\mathbb{E}\left[Y\left|X=z\right.\right]$ is given by:
\begin{equation}
\hat{f}_{\mathrm{NW}}(z):=\frac{\sum_{i=1}^{N}K_{h}(x_{i}-z)y_{i}}{\sum_{i=1}^{N}K_{h}(x_{i}-z)}\label{eq:localreg0}
\end{equation}

The estimator $\hat{f}_{\mathrm{NW}}(z)$ performs a kernel-weighted
local average of the response points $y_{i}$ that are such that their
corresponding inputs $x_{i}$ are close to the evaluation point $z$.
It can be described as a locally constant regression. More generally,
locally linear regressions can be performed:
\begin{equation}
\hat{f}_{\mathrm{L}}(z):=\min_{\alpha(z),\beta(z)}\sum_{i=1}^{N}K_{h}(x_{i}-z)\left[y_{i}-\alpha(z)-\beta(z)x_{i}\right]^{2}\label{eq:localreg1}
\end{equation}

In this case, a weighted linear regression is performed for each evaluation
point $z$ This formulation can be generalized to quadratic and higher-order
local polynomial regressions.

Discussions about the properties and performance of these classical
kernel smoothers \eqref{eq:localdens}-\eqref{eq:localreg0}-\eqref{eq:localreg1}
can be found in various textbooks, such as \citet{Loader99}, \citet{Hardle04},
\citet{Hastie09} and \citet{Scott14}. 

The well known computational problem with the implementation of the
kernel smoothers \eqref{eq:localdens}-\eqref{eq:localreg0}-\eqref{eq:localreg1}
is that their direct evaluation on a set of $M$ evaluation points
requires $\mathcal{O}(M\times N)$ operations. In particular, when
the evaluation points coincide with the input points $x_{1},x_{2},\ldots,x_{N}$,
a direct evaluation requires a quadratic $\mathcal{O}(N^{2})$ number
of operations. To cope with this computational limitation, several
approaches have been proposed over the years.

\textit{Data binning} consists in summarizing the input sample into
a set of equally spaced bins, so as to compute the kernel smoothers
more quickly on the binned data. This data preprocessing allows for
significant speedup, either by Fast Fourier Transform (\citet{Wand94},
\citet{Gramacki17}) or by direct computation, see \citet{Silverman82},
\citet{Scott85}, \citet{Fan94}, \citet{Turlachand96}, \citet{Bowman03}.

The \textit{fast sum updating} method is based on the sorting of the
input data and on a translation of the kernel from one evaluation
point to the next, updating only the input points which do not belong
to the intersection of the bandwidths of the two evaluation points,
see \citet{Gasser89}, \citet{Seifert94}, \citet{Fan94}, \citet{Werthenbach98},
\citet{Chen06}.

The \textit{Fast Gauss Transform}, also known as Fast Multipole Method,
is based on the expansion of the Gaussian kernel to disentangle the
input points from the evaluation points and speed up the evaluation
of the resulting sums, see \citet{Greengard91}, \citet{Greengard98},
\citet{Lambert99}, \citet{Yang03}, \citet{Morariu09}, \citet{Raykar10},
\citet{Sampath10}, \citet{Spivak10}.

The \textit{dual-tree} method is based on space partitioning trees
for both the input sample and the evaluation points. These tree structures
are then used to compute distances between input points and evaluation
points more quickly, see \citet{Gray01}, \citet{Gray03}, \citet{Lang05},
\citet{Lee06}, \citet{Ram09}, \citet{Curtin13}, \citet{Griebel13},
\citet{Lee14}.

Among all these methods, the fast sum updating is the only one which
is exact (no extra approximation is introduced) and whose speed is
independent of the input data, the kernel and the bandwidth. Its main
drawback is that the required sorting of the input points has mostly
limited this literature to the univariate case. \citet{Werthenbach98}
attempted to extend the method to the bivariate case, under strong
limitations, namely rectangular input sample, evaluation grid and
kernel support.

In this paper, we revisit the fast sum updating approach and extend
it to the general multivariate case. This extension requires a rectilinear
evaluation grid and kernels with box support, but has no restriction
on the input sample and can accommodate adaptive bandwidths. Moreover,
it maintains the desirable properties of the fast sum updating approach,
making it, so far, the only fast and exact algorithm for multivariate
kernel smoothing under general input sample and general bandwidth.

\section{Fast sum updating\label{sec:Fast-sum-updating}}

\subsection{Univariate case}

In this section, we recall the fast sum updating algorithm in the
univariate case. Let $(x_{1},y_{1})$, $(x_{2},y_{2})$,$\ldots$,$(x_{N},y_{N})$
be a sample of $N$ input (source) points $x_{i}$ and output points
$y_{i}$, and let $z_{1},z_{2},\ldots,z_{M}$ be a set of $M$ evaluation
(target) points. We first sort the input points and evaluation points:
$x_{1}\leq x_{2}\leq\ldots\leq x_{N}$ and $z_{1}\leq z_{2}\leq\ldots\leq z_{M}$.
In order to compute the kernel density estimator \eqref{eq:localdens},
the kernel regression \eqref{eq:localreg0} and the locally linear
regression \eqref{eq:localreg1} for every evaluation point $z_{j}$,
one needs to compute sums of the type
\begin{equation}
\mathbf{S}_{j}=\mathbf{S}_{j}^{p,q}:=\frac{1}{N}\sum_{i=1}^{N}K_{h}(x_{i}-z_{j})x_{i}^{p}y_{i}^{q}=\frac{1}{Nh}\sum_{i=1}^{N}K\left(\frac{x_{i}-z_{j}}{h}\right)x_{i}^{p}y_{i}^{q}\,,p=0,1,\,q=0,1\label{eq:sum_unidim}
\end{equation}
for every $j\in\{1,2,\ldots,M\}$. The direct, independent evaluation
of these sums would require $\mathcal{O}(N\times M)$ operations (a
sum of $N$ terms for each $j\in\{1,2,\ldots,M\}$). The idea of fast
sum updating is to use the information from the sum $\mathbf{S}_{j}$
to compute the next sum $\mathbf{S}_{j+1}$ without going through
all the $N$ input points again. We illustrate the idea with the Epanechnikov
(parabolic) kernel $K(u)=\frac{3}{4}(1-u^{2})\mathbbm{1}\{\left|u\right|\leq1\}$.
With this choice of kernel:
\begin{align}
 & \mathbf{S}_{j}^{p,q}=\frac{1}{Nh}\sum_{i=1}^{N}\frac{3}{4}\left(1-\left(\frac{x_{i}-z_{j}}{h}\right)^{2}\right)x_{i}^{p}y_{i}^{q}\mathbbm{1}\{z_{j}\!-\!h\leq x_{i}\leq z_{j}\!+\!h\}\nonumber \\
 & =\frac{1}{Nh}\frac{3}{4}\sum_{i=1}^{N}\left(1-\frac{z_{j}^{2}}{h^{2}}+2\frac{z_{j}}{h^{2}}x_{i}-\frac{1}{h^{2}}x_{i}^{2}\right)x_{i}^{p}y_{i}^{q}\mathbbm{1}\{z_{j}\!-\!h\leq x_{i}\leq z_{j}\!+\!h\}\nonumber \\
 & =\frac{3}{4Nh}\left\{ \!\left(\!1\!-\!\frac{z_{j}^{2}}{h^{2}}\!\right)\!\mathcal{S}^{p,q}([z_{j}\!-\!h,z_{j}\!+\!h])+2\frac{z_{j}}{h^{2}}\mathcal{S}^{p+1,q}([z_{j}\!-\!h,z_{j}\!+\!h])-\frac{1}{h^{2}}\mathcal{S}^{p+2,q}([z_{j}\!-\!h,z_{j}\!+\!h])\!\right\} \label{eq:parabolic-kernel-development}
\end{align}
where 
\begin{equation}
\mathcal{S}^{p,q}([L,R]):=\sum_{i=1}^{N}x_{i}^{p}y_{i}^{q}\mathbbm{1}\{L\leq x_{i}\leq R\}\label{eq:SpqLR}
\end{equation}
These sums $\mathcal{S}^{p,q}([z_{j}-h,z_{j}+h])$ can be evaluated
quickly from $j=1$ to $j=M$ as long as the input points $x_{i}$
and the evaluation points $z_{j}$ are sorted in increasing order.
Indeed,
\begin{align}
 & \mathcal{S}^{p,q}([z_{j+1}\!-\!h,z_{j+1}\!+\!h])=\sum_{i=1}^{N}x_{i}^{p}y_{i}^{q}\mathbbm{1}\{z_{j+1}\!-\!h\leq x_{i}\leq z_{j+1}\!+\!h\}\nonumber \\
 & =\sum_{i=1}^{N}x_{i}^{p}y_{i}^{q}\mathbbm{1}\{z_{j}\!-\!h\leq x_{i}\leq z_{j}\!+\!h\}\nonumber \\
 & -\sum_{i=1}^{N}x_{i}^{p}y_{i}^{q}\mathbbm{1}\{z_{j}\!-\!h\leq x_{i}<z_{j+1}\!-\!h\}+\sum_{i=1}^{N}x_{i}^{p}y_{i}^{q}\mathbbm{1}\{z_{j}\!+\!h<x_{i}\leq z_{j+1}\!+\!h\}\nonumber \\
 & =\mathcal{S}^{p,q}([z_{j}\!-\!h,z_{j}\!+\!h])-\mathcal{S}^{p,q}([z_{j}\!-\!h,z_{j+1}\!-\!h[)+\mathcal{S}^{p,q}(]z_{j}\!+\!h,z_{j+1}\!+\!h])\label{eq:1d_updating}
\end{align}
Therefore one can simply update the sum $\mathcal{S}^{p,q}([z_{j}-h,z_{j+1}+h])$
for the evaluation point $z_{j}$ to obtain the next sum $\mathcal{S}^{p,q}([z_{j+1}-h,z_{j+1}+h])$
for the next evaluation point $z_{j+1}$ by subtracting the terms
$x_{i}^{p}y_{i}^{q}$ for which $x_{i}$ lie between $z_{j}-h$ and
$z_{j+1}-h,$ and adding the terms $x_{i}^{p}y_{i}^{q}$ for which
$x_{i}$ lie between $z_{j}+h$ and $z_{j+1}+h$. This can be achieved
in a fast $\mathcal{O}(M+N)$ operations by going through the input
points $x_{i}$, stored in increasing order at a cost of $\mathcal{O}(N\log N)$
operations, and through the evaluation points $z_{j}$, stored in
increasing order at a cost of $\mathcal{O}(M\log M)$ operations.
Algorithm \ref{alg:fast1Dkreg} summarizes the whole procedure to
compute equations \eqref{eq:localdens}, \eqref{eq:localreg0} and
\eqref{eq:localreg1} in the case of the Epanechnikov kernel.

In the case of the Epanechnikov kernel, the expansion of the quadratic
term $\left(\frac{x_{i}-z_{j}}{h}\right)^{2}$ separates the sources
$x_{i}$ from the targets $z_{j}$ (equation \eqref{eq:parabolic-kernel-development}),
which makes the fast sum updating approach possible. Such a separation
occurs with other classical kernels as well, including the rectangular
kernel, the triangular kernel, the cosine kernel and the Silverman
kernel. Table \ref{tab:fast-sum-kernels} provides a list of ten kernels
for which fast sum updating can be implemented, and Appendix \ref{sec:fast-sum-kernels}
provides the detail of the updating formulas for these kernels. While
most of these kernels have finite support $[-1,1]$, some such as
the Laplacian kernel and Silverman kernel have infinite support. Not
every kernel admits such a separation between sources and targets,
the most prominent example being the Gaussian kernel $K(u)=\frac{1}{\sqrt{2\pi}}\exp(-u^{2}/2)$,
for which the cross term $\exp(x_{i}z_{j}/h)$ cannot be split between
one source term (depending on $i$ only) and one target term (depending
on $j$ only). Approximating the cross-term to obtain such a separation
is the path followed by the Fast Gauss Transform approach (\citet{Greengard91}).

While any kernel in Table \ref{tab:fast-sum-kernels} can be used
for fast sum updating, we choose to use for the rest of the paper
the Epanechnikov kernel $K(u)=\frac{3}{4}(1-u^{2})\mathbbm{1}\{\left|u\right|\leq1\}$
for two reasons: this popular kernel is optimal in the sense that
it minimizes the asymptotic mean integrated squared error (cf. \citet{Epanechnikov69}),
and it supports fast sum updating with adaptive bandwidth $h=h_{i}$
or $h=h_{j}$ (see Algorithm \ref{alg:fast1Dkreg}, Appendix \ref{sec:fast-sum-kernels}
and subsection \ref{subsec:knn-bandwidth}).\clearpage{}

\begin{algorithm2e}[H]
\DontPrintSemicolon
\SetAlgoLined
\vspace{1mm} 
 \KwIn{\\
  X: sorted vector of N inputs $X[1]\leq \ldots \leq X[N]$\\
  Y: vector of N outputs $Y[1], \ldots, Y[N]$\\
  Z: sorted vector of M evaluation points $Z[1]\leq \ldots\leq Z[M]$\\
  H: vector of M bandwidths $H[1], \ldots, H[M]$\\
\Comment*[l]{Z and H should be such that the vectors Z-H and Z+H are increasing}}  
  \vspace{1mm}
  iL = 1 \Comment*[l]{The indices $1\leq iL \leq iR \leq N$ will be such that the current}
  iR = 1 \Comment*[l]{bandwidth $[Z[m]-H[m],Z[m]+H[m]]$ contains the points $X[iL],X[iL+1],\ldots,X[iR]$}
  S[$p_1 , p_2$] = 0, $p_1=0,1,\ldots,4$, $p_2=0,1$ \Comment*[l]{Will contain the sum $\sum_{i=iL}^{iR} X[i]^{p_1}\times Y[i]^{p_2}$}
 \For{$m=1,...,M$}{
  \While{(iR$\leq$N) and (X[iR]$<$(Z[m]+H[m]))}{
   \vspace{0.5mm}
   S[$p_1 , p_2$] = S[$p_1 , p_2$] + X[iR]$^{p_1} \times$Y[iR]$^{p_2}$ , $p_1=0,1,\ldots,4,  p_2=0,1$\\
   iR = iR + 1\\
  }
  \While{(iL$\leq$N) and (X[iL]$<$(Z[m]-H[m]))}{
   \vspace{0.5mm}
   S[$p_1 , p_2$] = S[$p_1 , p_2$] $-$ X[iL]$^{p_1} \times$Y[iL]$^{p_2}$ , $p_1=0,1,\ldots,4,  p_2=0,1$\\
   iL = iL + 1\\
  }
  \Comment*[l]{Here S[$p_1,p_2$]=$\sum_{i=iL}^{iR} X[i]^{p_1} Y[i]^{p_2}$, which can be used to compute}
\vspace{0.5mm}
\Comment*[l]{SK[$p_{1},p_{2}$]=$\sum_{i=iL}^{iR}X[i]^{p_{1}}Y[i]^{p_{2}}\;K\!\left(Z[m],X[i]\right)$}
  \vspace{1mm}
  C0 = 1.0 $-$ Z[m]$^2$/H[m]$^2$; C1 = 2.0 $\times$ Z[m]/H[m]$^2$; C2 = 1/H[m]$^2$\\
  \vspace{0.5mm}
  SK[$p_1 , p_2$] = C0$\times$S[$p_1 , p_2$] + C1$\times$S[$p_1+1 , p_2$] $-$ C2$\times$S[$p_1+2 , p_2$]\\
  \vspace{0.5mm}
  D[m] = 0.75$\times$SK[0,0]/(H[m]$\times$N)\\  
  \vspace{0.5mm}
  R0[m] = SK[0,1]/SK[0,0]\\
  R1[m]=$\left[\begin{array}{cc} 1 & \mathrm{Z[m]}\end{array}\right]\left[\begin{array}{cc} \mathrm{SK[0,0]} & \mathrm{SK[1,0]}\\ \mathrm{SK[1,0]} & \mathrm{SK[2,0]} \end{array}\right]^{-1}\left[\begin{array}{c} \mathrm{SK[0,1]}\\ \mathrm{SK[1,1]} \end{array}\right]$

 }
return D, R0, R1\\ 
\vspace{1mm}
 \KwOut{\\
D[m]: kernel density estimate of X\\
R0[m]: locally constant regression of Y on X (kernel regression)\\
R1[m]: locally linear regression of Y on X\\
\Comment*[l]{The three estimates D[m], R0[m] and R1[m] are  evaluated at point Z[m] with bandwidth H[m] and Epanechnikov kernel, for each m=1,$\ldots$ ,M}}
\vspace{1mm}
 \caption{Fast univariate kernel smoothing\label{alg:fast1Dkreg}}
 \end{algorithm2e}

\begin{table}[H]
\bgroup 
\def\arraystretch{1.40}%
\begin{tabular}[t]{l>{\raggedright}m{10cm}}
\multicolumn{2}{l}{\textbf{Kernels compatible with fast sum updating}}\tabularnewline
\multirow{2}{*}{\includegraphics[height=18mm]{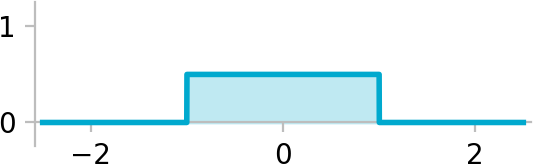}} & \textbf{Rectangular (uniform)} \vspace{0.1em}
\tabularnewline
 & $K(u)=\frac{1}{2}\mathbbm{1}\{\left|u\right|\leq1\}$ \vspace{1.5em}
\tabularnewline
\multirow{2}{*}{\includegraphics[height=18mm]{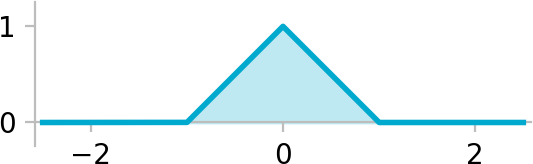}} & \textbf{Triangular}\vspace{0.1em}
\tabularnewline
 & $K(u)=(1-\left|u\right|)\mathbbm{1}\{\left|u\right|\leq1\}$\vspace{1.5em}
\tabularnewline
\multirow{2}{*}{\includegraphics[height=18mm]{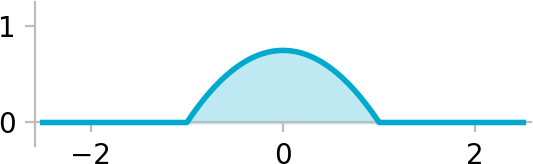}} & \textbf{Parabolic (Epanechnikov)}\vspace{0.1em}
\tabularnewline
 & $K(u)=\frac{3}{4}(1-u^{2})\mathbbm{1}\{\left|u\right|\leq1\}$\vspace{1.5em}
\tabularnewline
\multirow{2}{*}{\includegraphics[height=18mm]{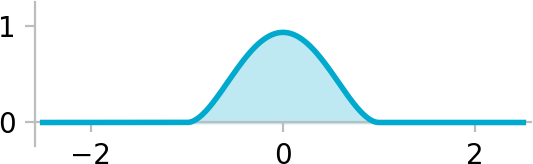}} & \textbf{Biweight (Quartic)}\vspace{0.1em}
\tabularnewline
 & $K(u)=\frac{15}{16}(1-u^{2})^{2}\mathbbm{1}\{\left|u\right|\leq1\}$\vspace{1.5em}
\tabularnewline
\multirow{2}{*}{\includegraphics[height=18mm]{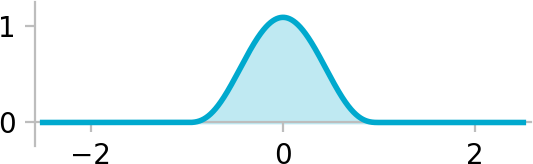}} & \textbf{Triweight}\vspace{0.1em}
\tabularnewline
 & $K(u)=\frac{35}{32}(1-u^{2})^{3}\mathbbm{1}\{\left|u\right|\leq1\}$\vspace{1.5em}
\tabularnewline
\multirow{2}{*}{\includegraphics[height=18mm]{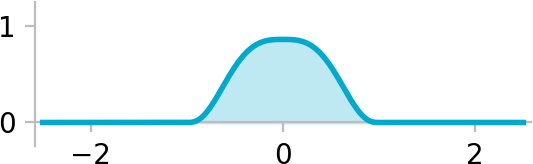}} & \textbf{Tricube}\vspace{0.1em}
\tabularnewline
 & $K(u)=\frac{70}{81}(1-\left|u\right|^{3})^{3}\mathbbm{1}\{\left|u\right|\leq1\}$\vspace{1.5em}
\tabularnewline
\multirow{2}{*}{\includegraphics[height=18mm]{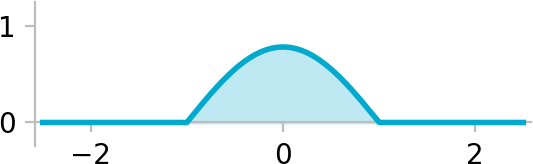}} & \textbf{Cosine}\vspace{0.1em}
\tabularnewline
 & $K(u)=\frac{\pi}{4}\cos\left(\frac{\pi}{2}u\right)\mathbbm{1}\{\left|u\right|\leq1\}$\vspace{1.5em}
\tabularnewline
\multirow{2}{*}{\includegraphics[height=18mm]{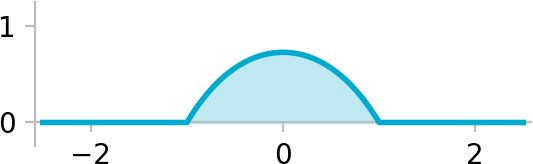}} & \textbf{Hyperbolic cosine}\vspace{0.1em}
\tabularnewline
 & $K(u)=\frac{1}{4-2\frac{\sinh(\log(2+\!\sqrt{3}))}{\log(2+\!\sqrt{3})}}\left\{ 2-\cosh(\log(2+\!\sqrt{3})u)\right\} \mathbbm{1}\{\left|u\right|\!\leq\!1\}$\vspace{0.75em}
\tabularnewline
\multirow{2}{*}{\includegraphics[height=18mm]{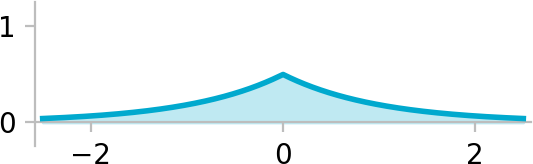}} & \textbf{Laplacian}\vspace{0.1em}
\tabularnewline
 & $K(u)=\frac{1}{2}\exp(-\left|u\right|)$\vspace{1.5em}
\tabularnewline
\multirow{2}{*}{\includegraphics[height=18mm]{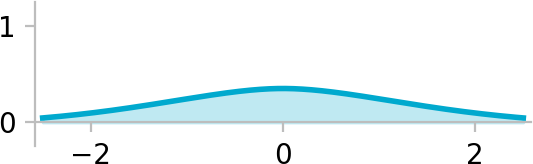}} & \textbf{Silverman}\vspace{0.1em}
\tabularnewline
 & $K(u)=\frac{1}{2}\exp\left(-\frac{\left|u\right|}{\sqrt{2}}\right)\sin\left(\frac{\left|u\right|}{\sqrt{2}}+\frac{\pi}{4}\right)$\vspace{1.5em}
\tabularnewline
\end{tabular}\egroup \caption{Kernels compatible with fast sum updating\label{tab:fast-sum-kernels}}
\end{table}

\clearpage{}

\subsection{Numerical stability\label{subsec:Numerical-stability}}

In \citet{Seifert94}, the direct fast sum updating approach described
in Algorithm \ref{alg:fast1Dkreg} was discarded for numerical stability
reasons. With floating-point arithmetic, the difference $(x+y)-x$
is in general equal to $y\pm\varepsilon$, where $\varepsilon$ corresponds
to the floating point rounding error. In addition, the greater the
scale difference between two floating numbers $x$ and $y$, the greater
the rounding error when computing $x+y$. Consequently, adding and
subtracting $N$ numbers in sequence has a worst-case rounding error
that grows proportional to $N$. 

In this paper, we argue that the advances in floating-point accuracy
and stable floating-point summation in the past decades have made
the direct fast sum updating approach viable and immune to numerical
error. In addition to simple precautions such as normalization of
input data and use of accurate floating-point formats such as quadruple-precision
floating-point, a long list of stable summation algorithms have been
proposed in the past fifty years, see among others \citet{Moller65},
\citet{Kahan65}, \citet{Linnainmaa74}, \citet{Priest91} \citet{Higham93},
\citet{Demmel03}, \citet{McNamee04} and \citet{Boldo17}. The usual
idea is to keep track of the current amount of floating-point rounding
error, and to propagate it when adding new terms in the sum. Recently,
a number of exact summation algorithms have been proposed, see \citet{Rump08},
\citet{Pan09}, \citet{Zhu10} and \citet{Neal15}. These algorithms
are exact in the sense that the final result is the closest floating-point
number, within the precision of the chosen floating-point format,
to the exact mathematical sum of the inputs. Importantly, the computational
complexity of exact summation remains linear in the number $N$ of
data points to sum. For example, for the recent \citet{Neal15}, exact
summation is less than a factor two slower than naive summation.

To sum up, with little modification, fast sum updating algorithms
such as Algorithm \ref{alg:fast1Dkreg} and its multivariate version
\ref{alg:fastdDkreg} can be made completely immune to numerical instability.
As a simple illustration, Algorithm \ref{alg:fast1DkregMK} in Appendix
shows how to combine Algorithm \ref{alg:fastdDkreg} with the stable
M\o ller-Kahan summation algorithm (\citet{Moller65}). For simplicity
and clarity, the fast sum updating algorithms presented in this paper
omit the stabilisation components. All of them can be implemented
with perfect numerical stability using the stable summation algorithms
mentioned in this subsection. 

\subsection{Multivariate case}

We now turn to the multivariate case. Let $d$ be the dimension of
the inputs. We consider again a sample $(x_{1},y_{1}),(x_{2},y_{2}),\ldots,(x_{N},y_{N})$
of $N$ input points $x_{i}$ and output points $y_{i}$, where the
input points are now multivariate:
\[
x_{i}=\left(x_{1,i},x_{2,i},\ldots,x_{d,i}\right)\,,\,i\in\{1,2,\ldots,N\}
\]

\subsubsection{Multivariate kernel smoothers\label{subsec:Multivariate-kernel-smoothers}}

The kernel smoothers \eqref{eq:localdens}, \eqref{eq:localreg0}
and \eqref{eq:localreg1} can be extended to the multivariate case.
A general form for a multivariate kernel is $K_{d,H}(u)=\left|H\right|^{-1/2}K_{d}(H^{-1/2}u)$,
where $u=(u_{1},u_{2},\ldots,u_{d})\in\mathbb{R}^{d}$ and where $H$
is a symmetric positive definite $d\times d$ bandwidth matrix (see
\citet{Wand95} for example). The eigenvalue decomposition of $H$
yields $H=R\Delta^{2}R^{\top}$ where $R$ is a rotation matrix and
$\Delta=\mathrm{diag}(h)$ is a diagonal matrix with strictly positive
diagonal elements $h=(h_{1},h_{2},\ldots,h_{d})\in\mathbb{R}^{d}$.
Therefore, without loss of generality, one can focus on the diagonal
bandwidth case $K_{d,h}(u)=\frac{1}{\Pi_{k=1}^{d}h_{k}}K_{d}(\frac{u_{1}}{h_{1}},\frac{u_{2}}{h_{2}},\ldots,\frac{u_{d}}{h_{d}})$
after a rotation of the input points $x_{i}$ and the evaluation points
$z_{j}$ using $R$. Subsection \ref{subsec:grid-shape} will discuss
the choice of data rotation and subsection \ref{subsubsec:multivariate-kernel}
will discuss the possible choices of multivariate kernels $K_{d}$
compatible with fast sum updating. One can show (cf. Appendix \ref{sec:Multivariate-kernel-smoothers})
that the computation of the multivariate version of the kernels smoothers
\eqref{eq:localdens}, \eqref{eq:localreg0} and \eqref{eq:localreg1}
boils down to the computation of the following sums:
\begin{eqnarray}
\mathbf{S}_{j} & = & \mathbf{S}_{k_{1},k_{2},j}^{p_{1},p_{2},q}:=\frac{1}{N}\sum_{i=1}^{N}K_{d,h}(x_{i}-z_{j})x_{k_{1},i}^{p_{1}}x_{k_{2},i}^{p_{2}}y_{i}^{q}\nonumber \\
 & = & \frac{1}{N\Pi_{k=1}^{d}h_{k}}\sum_{i=1}^{N}K_{d}\left(\frac{x_{1,i}-z_{1,j}}{h_{1}},\frac{x_{2,i}-z_{2,j}}{h_{2}},\ldots,\frac{x_{d,i}-z_{d,j}}{h_{d}}\right)x_{k_{1},i}^{p_{1}}x_{k_{2},i}^{p_{2}}y_{i}^{q}\label{eq:sum_multidim}
\end{eqnarray}
for each evaluation point $z_{j}=(z_{1,j},z_{2,j},\ldots,z_{d,j})\in\mathbb{R}^{d}$,
$j\in\{1,2,\ldots,M\}$, for powers $p_{1},p_{2},q=0,1$ and for dimension
indices $k_{1},k_{2}=1,2,\ldots,d$.

Before expanding the sum \eqref{eq:sum_multidim} as was done in \eqref{eq:parabolic-kernel-development}
in the univariate case, we first introduce the two conditions required
for fast multivariate sum updating (subsection \ref{subsec:Conditions})
and then discuss the choice of multivariate kernel (subsection \ref{subsubsec:multivariate-kernel}).

\subsubsection{Conditions\label{subsec:Conditions}}

In order to extend the fast sum updating algorithm to the multivariate
case, we require the following two conditions:
\begin{condition}
\label{cond:=00005BEvaluation-grid=00005D}{[}Evaluation grid{]} We
require the evaluation grid to be rectilinear, i.e., the $M$ evaluation
points $z_{1},z_{2},\ldots,z_{M}$ lie on a regular grid with possibly
non-uniform mesh, of dimension $M_{1}\times M_{2}\times\ldots\times M_{d}=M$:
\[
\left\{ (z_{1,j_{1}},z_{2,j_{2}},\ldots,z_{d,j_{d}})\in\mathbb{R}^{d},\,j_{k}\in\{1,2,\ldots,M_{k}\},\,k\in\{1,2,\ldots,d\}\right\} 
\]
Figure \ref{fig:evaluation grid} on page \pageref{fig:evaluation grid}
provides two examples of rectilinear evaluation grids in the bivariate
case.
\end{condition}
\begin{condition}
\label{cond:=00005BKernel-support=00005D}{[}Kernel support{]} We
allow the bandwidths to vary with the evaluation points (balloon estimators,
see subsection \ref{subsec:knn-bandwidth}) but require them to follow
the shape of the evaluation grid. In other words, each evaluation
point $z_{j}=\left(z_{1,j_{1}},z_{2,j_{2}},\ldots,z_{d,j_{d}}\right)$
is associated with its own bandwidth $h_{j}=\left(h_{1,j_{1}},h_{2,j_{2}},\ldots,h_{d,j_{d}}\right)$.
For kernels with finite support (first eight kernels in Table \ref{tab:fast-sum-kernels}),
this means that the kernel support must be a hyperrectangle, i.e.
the box 
\begin{align*}
 & \prod_{k=1}^{d}\left[z_{k,j_{k}}-h_{k,j_{k}},z_{k,j_{k}}+h_{k,j_{k}}\right]\\
 & :=\left\{ \left(u_{1},u_{2},\ldots,u_{d}\right)\in\mathbb{R}^{d}\left|u_{k}\in\left[z_{k,j_{k}}-h_{k,j_{k}},z_{k,j_{k}}+h_{k,j_{k}}\right],k=1,2,\ldots,d\right.\right\} 
\end{align*}
where $j_{k}\in\{1,2,\ldots,M_{k}\},\,k\in\{1,2,\ldots d\}$. 
\end{condition}
The reason for these two conditions will become clear after the description
of the multivariate sweeping algorithm for multivariate sum updating.
In the rest of this section, we assume these two conditions are satisfied.
The next subsection discusses the choice of multivariate kernel, and
show that two simple types of multivariate kernels satisfy Condition
\ref{cond:=00005BKernel-support=00005D}: product kernels (equation
\eqref{eq:kernel_prod}) and average kernels (equation \eqref{eq:kernel_mean}).

\subsubsection{Multivariate kernel\label{subsubsec:multivariate-kernel}}

\begin{figure}[h]
\begin{centering}
\includegraphics[width=0.75\paperwidth]{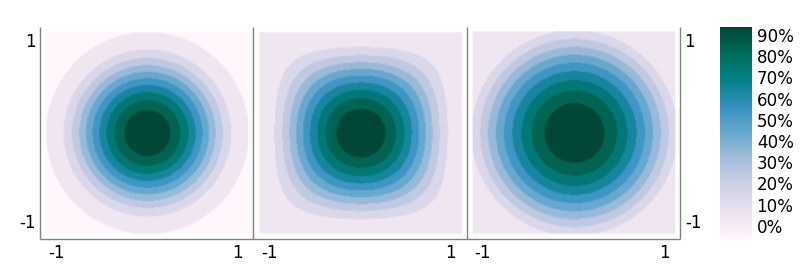}
\par\end{centering}
\caption{Bivariate parabolic kernels\label{fig:Bivariate-parabolic-kernels}}
\end{figure}

To extend the definitions of the smoothing kernels \eqref{eq:localdens},
\eqref{eq:localreg0} and \eqref{eq:localreg1} to the multivariate
case, one needs kernel functions defined in a multivariate setting.
There exists different ways to extend a univariate kernel to the multivariate
case, see \citet{Hardle00} for example. As an illustration, Figure
\ref{fig:Bivariate-parabolic-kernels} displays three different ways
to extend the Epanechnikov kernel $K_{1}(u)=\frac{3}{4}\left(1-u^{2}\right)$
to the multivariate (bivariate) case.

The left-side kernel in Figure \ref{fig:Bivariate-parabolic-kernels}
corresponds to the \textit{spherical} or \textit{radially symmetric}
kernel:
\begin{equation}
K_{d}^{S}\left(u_{1},\ldots,u_{d}\right)=\frac{\Gamma\left(2+\frac{d}{2}\right)}{\pi^{\frac{d}{2}}}\left(1-\left\Vert u\right\Vert ^{2}\right)\mathbbm{1}\{\left\Vert u\right\Vert \leq1\}\label{eq:kernel_norm}
\end{equation}
for which the norm of the vector $u$ is used as an input in the univariate
kernel (with a proper normalization constant, see \citet{Fukunaga75}).
This multivariate kernel is the most efficient in terms of asymptotic
mean integrated squared error (see \citet{Wand95} for example). Unfortunately,
this kernel is not compatible with fast sum updating, as its support
is a hypersphere, while Condition \ref{cond:=00005BKernel-support=00005D}
requires a hyperrectangle support. The middle kernel in Figure \ref{fig:Bivariate-parabolic-kernels}
corresponds to the \textit{multiplicative} or \textit{product} kernel:
\begin{equation}
K_{d}^{P}\left(u_{1},\ldots,u_{d}\right)=\prod_{k=1}^{d}K_{1}\left(u_{k}\right)=\left(\frac{3}{4}\right)^{d}\prod_{k=1}^{d}\left\{ \left(1-u_{k}^{2}\right)\mathbbm{1}\{\left|u_{k}\right|\leq1\}\right\} \label{eq:kernel_prod}
\end{equation}
obtained by multiplying univariate kernels. Its support is a hyperrectangle.
Finally, the right-side kernel in Figure \ref{fig:Bivariate-parabolic-kernels}
corresponds to the \textit{additive} or \textit{arithmetic average}
kernel:
\begin{equation}
K_{d}^{A}\left(u_{1},\ldots,u_{d}\right)=\frac{1}{d2^{d-1}}\sum_{k=1}^{d}K_{1}(u_{k})\prod_{\substack{k_{0}=1\\
k_{0}\neq k
}
}^{d}\mathbbm{1}\{\left|u_{k_{0}}\right|<1\}=\frac{3}{d2^{d+1}}\sum_{k=1}^{d}\left(1-u_{k}^{2}\right)\prod_{k_{0}=1}^{d}\mathbbm{1}\{\left|u_{k_{0}}\right|<1\}\label{eq:kernel_mean}
\end{equation}
which is obtained by averaging univariate kernels, and is another
general way of producing multivariate kernels. The support of this
kernel is also a hyperrectangle. As Condition \ref{cond:=00005BKernel-support=00005D}
rules out the spherical kernel \eqref{eq:kernel_norm}, we have to
make a choice between the product kernel \eqref{eq:kernel_prod} and
the average kernel \eqref{eq:kernel_mean}. When it comes to choosing
a kernel, the following quote from \citet{Silverman82} summarizes
the general consensus in the literature: \textit{``Both theory and
practice suggest that the choice of kernel is not crucial to the statistical
performance of the method and therefore it is quite reasonable to
choose a kernel for computational efficiency''}. In our context, this
observation means that the average kernel \eqref{eq:kernel_mean}
is to be preferred over the product kernel \eqref{eq:kernel_prod}
for its greater computational efficiency. Indeed, while average kernels
are not as efficient\footnote{The efficiency $\mathrm{eff}(K)$ of a kernel $K$ is defined as the
ratio $R(K^{S})\mu_{2}^{d/2}(K^{S})/(R(K)\mu_{2}^{d/2}(K))$ where
$R(K):=\int\cdots\int K^{2}(u_{1},\ldots,u_{d})du_{1}\ldots du_{d}$,
$\mu_{2}(K):=\int\cdots\int u_{1}^{2}K(u_{1},\ldots,u_{d})du_{1}\ldots du_{d}$
and $K^{S}$ is the spherical kernel \eqref{eq:kernel_norm}, see
\citet{Wand95}. The speedup of $K^{A}$ over $K^{P}$ to achieve
the same accuracy (Table \ref{tab:additive_kernel}), is defined as
$3^{d}\mathrm{eff}(K^{P})/((2d+1)\mathrm{eff}(K^{A}))=(18/5)^{d}(3d/(5d-2))^{d/2}5d/((2d+1)(5d+1))$.} as product kernels (see Table \ref{tab:additive_kernel}) they contain
much fewer sums to track down for the fast sum updating algorithm
(after expanding the squared terms $(x_{k,i}-z_{k,j})^{2}/h_{k}^{2}$,
the sum \eqref{eq:sum_multidim} is composed of $3^{d}$ different
sums over $i=1,\ldots,N$ for the product kernel \eqref{eq:kernel_prod},
compared to only $2d+1$ sums for the average kernel \eqref{eq:kernel_mean}).
In the end, to achieve the same accuracy, the average kernel \eqref{eq:kernel_mean}
is vastly faster than the product kernel \eqref{eq:kernel_prod} when
using the fast sum updating approach (around 80\% faster for bivariate
problems, more than 18 times faster for five-dimensional problems,
see Table \ref{tab:additive_kernel}). For this reason, we henceforth
use the average multivariate kernel \eqref{eq:kernel_mean} in the
rest of the paper.

\begin{table}[h] 
\begin{centering} 
\begin{tabular}{lllll} 
\toprule 
$\!\!\!$dimension & 2D & 3D & 4D & 5D\tabularnewline 
\midrule 
$\!\!\!$$K^{P}$ efficiency & 98.2\% & 95.3\% & 91.6\% & 87.4\%\tabularnewline 
$\!\!\!$$K^{A}$ efficiency & 96.5\% & 88.9\% & 80.4\% & 71.8\% \tabularnewline \midrule 
$\!\!\!$$K^{P}$ number of sums & 9 & 27 & 81 & 243\tabularnewline 
$\!\!\!$$K^{A}$ number of sums & 5 & 7 & 9 & 11\tabularnewline \midrule 
$\!\!\!$speedup factor of $K^{A}$ over $K^{P}$ & 1.8 & 3.6 & 7.9 & 18.2\tabularnewline 
\bottomrule 
\end{tabular} 
\par\end{centering} 
\centering{} \caption{product kernel $K^{P}$ vs. average kernel $K^{A}$\label{tab:additive_kernel}} 
\end{table}

\subsubsection{Kernel expansion}

Using the multivariate kernel \eqref{eq:kernel_mean}, one can expand
the sum \eqref{eq:sum_multidim} as follows:

\begin{align}
 & \mathbf{S}_{j}:=\mathbf{S}_{k_{1},k_{2},j}^{p_{1},p_{2},q}=\frac{1}{N\prod_{k=1}^{d}h_{k}}\sum_{i=1}^{N}K_{d}\left(\frac{x_{1,i}-z_{1,j}}{h_{1}},\frac{x_{2,i}-z_{2,j}}{h_{2}},\ldots,\frac{x_{d,i}-z_{d,j}}{h_{d}}\right)x_{k_{1},i}^{p_{1}}x_{k_{2},i}^{p_{2}}y_{i}^{q}\nonumber \\
 & =\frac{3}{d2^{d+1}N\prod_{k=1}^{d}h_{k}}\sum_{i=1}^{N}\sum_{k=1}^{d}\left(1-\frac{(x_{k,i}-z_{k,j})^{2}}{h_{k}^{2}}\right)x_{k_{1},i}^{p_{1}}x_{k_{2},i}^{p_{2}}y_{i}^{q}\prod_{k_{0}=1}^{d}\mathbbm{1}\{\left|x_{k_{0},i}-z_{k_{0},j}\right|\leq1\}\nonumber \\
 & =\frac{3}{d2^{d+1}N\prod_{k=1}^{d}h_{k}}\sum_{k=1}^{d}\sum_{i=1}^{N}\left(1-\frac{z_{k,j}^{2}}{h_{k}^{2}}+2\frac{z_{k,j}}{h_{k}^{2}}x_{k,i}-\frac{1}{h_{k}^{2}}x_{k,i}^{2}\right)x_{k_{1},i}^{p_{1}}x_{k_{2},i}^{p_{2}}y_{i}^{q}\prod_{k_{0}=1}^{d}\mathbbm{1}\{\left|x_{k_{0},i}-z_{k_{0},j}\right|\leq1\}\nonumber \\
 & =\frac{3}{d2^{d+1}N\prod_{k=1}^{d}h_{k}}\sum_{k=1}^{d}\left\{ \left(1-\frac{z_{k,j}^{2}}{h_{k}^{2}}\right)\mathcal{S}_{[k,k_{1},k_{2}]}^{[0,p_{1},p_{2}],q}([z_{j}-h_{j},z_{j}+h_{j}])+\right.\nonumber \\
 & =\left.2\frac{z_{k,j}}{h_{k}^{2}}\mathcal{S}_{[k,k_{1},k_{2}]}^{[1,p_{1},p_{2}],q}([z_{j}-h_{j},z_{j}+h_{j}])-\frac{1}{h_{k}^{2}}\mathcal{S}_{[k,k_{1},k_{2}]}^{[2,p_{1},p_{2}],q}([z_{j}-h_{j},z_{j}+h_{j}])\right\} \label{eq:parabolic-kernel-development-multivariate}
\end{align}

where for any hyperrectangle $[\mathbf{L},\mathbf{R}]:=\left[L_{1},R_{1}\right]\times\left[L_{2},R_{2}\right]\times\ldots\times\left[L_{d},R_{d}\right]\subseteq\mathbb{R}^{d}$:
\begin{equation}
\mathcal{S}^{\mathrm{idx}}([\mathbf{L},\mathbf{R}]):=\mathcal{S}_{\mathbf{k}}^{\mathbf{p},q}([\mathbf{L},\mathbf{R}]):=\sum_{i=1}^{N}\left(\prod_{l=1}^{3}(x_{k_{l},i})^{p_{l}}\right)y_{i}^{q}\prod_{k_{0}=1}^{d}\mathbbm{1}\{L_{k_{0}}\leq x_{k_{0},i}\leq R_{k_{0}}\}\label{eq:SpqLRd}
\end{equation}

for powers $\mathbf{p}:=(p_{1},p_{2},p_{3})\in\mathbb{N}^{3}$, $q\in\mathbb{N}$
and indices $\mathbf{k}:=(k_{1},k_{2},k_{3})\in\{1,2,\ldots,d\}^{3}$,
and where $[z_{j}-h_{j},z_{j}+h_{j}]:=\left[z_{1,j}-h_{1,j},z_{1,j}+h_{1,j}\right]\times\ldots\times\left[z_{d,j}-h_{d,j},z_{d,j}+h_{d,j}\right]$.
To simplify notations, we make use of the multi-index $\mathrm{idx}:=(\mathbf{p},q,\mathbf{k})$.

To sum up what has been obtained so far, computing multivariate kernel
smoothers (kernel density estimation, kernel regression, locally linear
regression) boils down to computing sums of the type \eqref{eq:SpqLRd}
on hyperrectangles of the type $[z_{j}-h_{j},z_{j}+h_{j}]$ for every
evaluation point $j\in\{1,2,\ldots,M\}$. In the univariate case,
these sums could be computed efficiently by sorting the input points
$x_{i}$, $i\in\{1,2,\ldots,N\}$ and updating the sums from one evaluation
point to the next (equation \eqref{eq:1d_updating}). Our goal is
now to set up a similar efficient fast sum updating algorithm for
the multivariate sums \eqref{eq:SpqLRd}. To do so, we first partition
the input data into a multivariate rectilinear grid (subsection \ref{subsec:Data-partition}),
by taking advantage of the fact that the evaluation grid is rectilinear
(Condition \ref{cond:=00005BEvaluation-grid=00005D}) and that the
support of the kernels has a hyperrectangle shape (Condition \ref{cond:=00005BKernel-support=00005D}).
Then, we set up a fast sweeping algorithm using the sums on each hyperrectangle
of the partition as the unit blocks to be added and removed (subsection
\ref{subsec:Fast-multivariate-sweeping}), unlike the univariate case
where the input points themselves were being added and removed iteratively.
Finally, the computational speed of this new algorithm is discussed
in subsection \ref{subsec:Complexity}.

\subsubsection{Data partition\label{subsec:Data-partition}}

\begin{figure}[h]
\begin{centering}
\includegraphics[width=0.75\paperwidth]{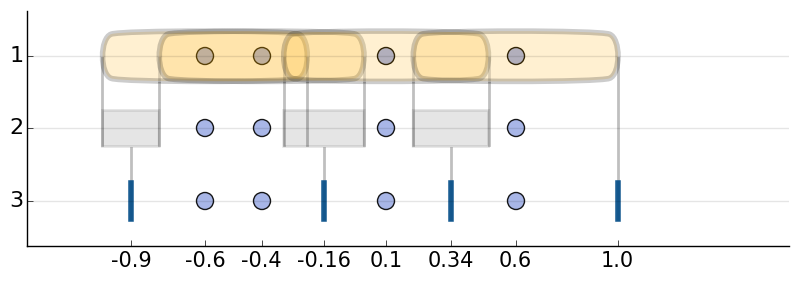}
\par\end{centering}
\caption{From bandwidths to partition (1D)\label{fig:1D_partition}}
\end{figure}

The first stage of the multivariate fast sum updating algorithm is
to partition the sample of input points into boxes. To do so, define
the sorted lists 
\[
\tilde{\mathcal{G}}_{k}=\left\{ \tilde{g}_{k,1},\tilde{g}_{k,2},\ldots,\tilde{g}_{k,2M_{k}}\right\} :=\mathrm{sort}\!\left(\left\{ z_{k,j_{k}}-h_{k,j_{k}}\right\} _{j_{k}\in\{1,2,\ldots,M_{k}\}}\bigcup\left\{ z_{k,j_{k}}+h_{k,j_{k}}\right\} _{j_{k}\in\{1,2,\ldots,M_{k}\}}\right)
\]
 in each dimension $k\in\{1,2,\ldots,d\}$, and define the partition
intervals $\tilde{I}_{k,l}:=\left[\tilde{g}_{k,l},\tilde{g}_{k,l+1}\right]$
for $l\in\left\{ 1,2,\ldots,2M_{k}-1\right\} $. The second row of
Figure \ref{fig:1D_partition} illustrates this partition on a set
of $4$ points, where for simplicity the evaluation points are the
same as the input points. By definition of $\tilde{\mathcal{G}}_{k}$,
all the bandwidths edges $z_{k,j_{k}}-h_{k,j_{k}}$ and $z_{k,j_{k}}+h_{k,j_{k}}$,
$j_{k}\in\{1,2,\ldots,M_{k}\}$, belong to $\tilde{\mathcal{G}}_{k}$.
Therefore, there exists some indices $\tilde{L}_{k,j_{k}}$ and $\tilde{R}_{k,j_{k}}$
such that 
\[
[z_{k,j_{k}}-h_{k,j_{k}},z_{k,j_{k}}+h_{k,j_{k}}]=[\tilde{g}_{k,\tilde{L}_{k,j_{k}}},\tilde{g}_{k,\tilde{R}_{k,j_{k}}\!+1}]=\bigcup_{l_{k}\in\{\tilde{L}_{k,j_{k}}\!,\ldots,\tilde{R}_{k,j_{k}}\!\}}\!\!\tilde{I}_{k,l_{k}}\,.
\]
From there, for any evaluation point $z_{j}=\left(z_{1,j_{1}},z_{2,j_{2}},\ldots,z_{d,j_{d}}\right)\in\mathbb{R}^{d}$,
the box $[z_{j}-h_{j},z_{j}+h_{j}]\subset\mathbb{R}^{d}$ can be decomposed
into a union of smaller boxes:
\begin{align}
[z_{j}-h_{j},z_{j}+h_{j}] & =\left[z_{1,j_{1}}-h_{1,j_{1}},z_{1,j_{1}}+h_{1,j_{1}}\right]\times\ldots\times\left[z_{d,j_{d}}-h_{d,j_{d}},z_{d,j_{d}}+h_{d,j_{d}}\right]\nonumber \\
 & =\left[\tilde{g}_{1,\tilde{L}_{1,j_{1}}},\tilde{g}_{1,\tilde{R}_{1,j_{1}}\!+1}\right]\times\ldots\times\left[\tilde{g}_{d,\tilde{L}_{d,j_{d}}},\tilde{g}_{d,\tilde{R}_{d,j_{d}}\!+1}\right]\nonumber \\
 & =\bigcup_{(l_{1}\!,\ldots,l_{d})\in\{\tilde{L}_{1,j_{1}}\!,\ldots,\tilde{R}_{1,j_{1}}\!\}\!\times\ldots\times\!\{\tilde{L}_{d,j_{d}}\!,\ldots,\tilde{R}_{d,j_{d}}\!\}}\!\!\!\tilde{I}_{1,l_{1}}\!\times\ldots\times\tilde{I}_{d,l_{d}}\label{eq:pavement}
\end{align}
In other words, the set of boxes $\tilde{I}_{1,l_{1}}\times\tilde{I}_{2,l_{2}}\times\ldots\times\tilde{I}_{d,l_{d}}$
s.t. $l_{k}\in\{\tilde{L}_{k,j_{k}},\tilde{L}_{k,j_{k}}+1,\ldots,\tilde{R}_{k,j_{k}}\}$
in each dimension $k\in\left\{ 1,2,\ldots,d\right\} $ forms a partition
of the box $[z_{j}-h_{j},z_{j}+h_{j}]$. Consequently, the sum \eqref{eq:SpqLRd}
evaluated on the box {[}$z_{j}-h_{j},z_{j}+h_{j}]$ can be decomposed
as follows:
\begin{equation}
\mathcal{S}^{\mathrm{idx}}([z_{j}-h_{j},z_{j}+h_{j}])=\!\sum_{(l_{1}\!,\ldots,l_{d})\in\{\tilde{L}_{1,j_{1}}\!,\ldots,\tilde{R}_{1,j_{1}}\!\}\!\times\ldots\times\!\{\tilde{L}_{d,j_{d}}\!,\ldots,\tilde{R}_{d,j_{d}}\!\}}\!\!\!\mathcal{S}^{\mathrm{idx}}\!\left(\tilde{I}_{1,l_{1}}\times\ldots\times\tilde{I}_{d,l_{d}}\right)\label{eq:sum_partition_full}
\end{equation}
where we assume without loss of generality that the bandwidth grid
$h_{j}\!=\!\left(h_{1,j_{1}},\!h_{2,j_{2}},\!\ldots\!,\!h_{d,j_{d}}\right)$,
$j_{k}\in\{1,2,\ldots,M_{k}\},\,k\in\{1,2,\ldots d\}$ is such that
the list $\tilde{\mathcal{G}}_{k}$ does not contain any input $x_{k,i}$,
$i\in\left\{ 1,2,\ldots,N\right\} $ (as such boundary points would
be counted twice in the right-hand side of \eqref{eq:sum_partition_full}).
This simple condition is easy to satisfy, as shown by the adaptive
bandwidth example provided in subsection \ref{subsec:knn-bandwidth}.

The sum decomposition \eqref{eq:sum_partition_full} is the cornerstone
of the fast multivariate sum updating algorithm, but before going
further, one can simplify the partitions $\tilde{\mathcal{G}}_{k}$,
$k\in\left\{ 1,2,\ldots,d\right\} $ while maintaining a sum decomposition
of the type \eqref{eq:sum_partition_full}. Indeed, in general some
intervals $\tilde{I}_{k,l}$ might be empty (i.e. they might not contain
any input point $x_{k,i}$, cf. the grey intervals on the second row
of Figure \ref{fig:1D_partition}). To avoid keeping track of sums
$\mathcal{S}^{\mathrm{idx}}$ on boxes known to be empty, one can
trim the partitions $\tilde{\mathcal{G}}_{k}$ by replacing each succession
of empty intervals by one new partition threshold. For example, if
$\tilde{I}_{k,l}=[\tilde{g}_{k,l},\tilde{g}_{k,l+1}]$ is empty, one
can remove the two points $\tilde{g}_{k,l}$ and $\tilde{g}_{k,l+1}$
and replace them by, for example, $(\tilde{g}_{k,l}+\tilde{g}_{k,l+1})/2$
(cf. the final partition on the third row of Figure \ref{fig:1D_partition}).
Denote by $\mathcal{G}_{k}=\left\{ g_{k,1},g_{k,2},\ldots,g_{k,m_{k}}\right\} $
the sorted simplified list, where $2\leq m_{k}\leq2M_{k}$, $k\in\left\{ 1,2,\ldots,d\right\} $,
and $m:=\prod_{k=1}^{d}m_{k}\leq2^{d}M$. Define the new partition
intervals $I_{k,l}:=\left[g_{k,l},g_{k,l+1}\right]$, $l\in\left\{ 1,2,\ldots,m_{k}-1\right\} $.
Because the trimming from $\tilde{\mathcal{G}}_{k}$ to $\mathcal{G}_{k}$
only affects the empty intervals, the following still holds:
\begin{lem}
\label{lem:sum_decomposition}For any evaluation point $z_{j}=\left(z_{1,j_{1}},z_{2,j_{2}},\ldots,z_{d,j_{d}}\right)\in\mathbb{R}^{d}$,
$j_{k}\in\{1,2,\ldots,M_{k}\},\,k\in\{1,2,\ldots d\}$ , there exists
indices $\left(L_{1,j_{1}},L_{2,j_{2}},\ldots,L_{d,j_{d}}\right)$
and $\left(R_{1,j_{1}},R_{2,j_{2}},\ldots,R_{d,j_{d}}\right)$, where
$L_{k,j_{k}}\in\{1,2,\ldots,m_{k}-1\}$ and $R_{k,j_{k}}\in\{1,2,\ldots,m_{k}-1\}$
with $L_{k,j_{k}}\leq R_{k,j_{k}}$, $k\in\{1,2,\ldots d\}$, such
that
\begin{equation}
\mathcal{S}^{\mathrm{idx}}([z_{j}-h_{j},z_{j}+h_{j}])=\!\sum_{(l_{1}\!,\ldots,l_{d})\in\{L_{1,j_{1}}\!,\ldots,R_{1,j_{1}}\!\}\!\times\ldots\times\!\{L_{d,j_{d}}\!,\ldots,R_{d,j_{d}}\!\}}\!\!\!\mathcal{S}^{\mathrm{idx}}\!\left(I_{1,l_{1}}\times\ldots\times I_{d,l_{d}}\right)\label{eq:sum_partition_slim}
\end{equation}

For later use, we introduce the compact notation $\mathcal{S}_{l_{1},l_{2},\ldots,l_{d}}^{\mathrm{idx}}\hspace{-0.1em}:=\mathcal{S}^{\mathrm{idx}}(I_{1,l_{1}}\hspace{-0.1em}\!\times\!\ldots\!\times\hspace{-0.1em}I_{d,l_{d}})$.
Recalling equation \eqref{eq:SpqLRd}, the sum $\mathcal{S}_{l_{1},l_{2},\ldots,l_{d}}^{\mathrm{idx}}$
corresponds to the sum of the polynomials $(\prod_{l=1}^{3}(x_{k_{l},i})^{p_{l}})y_{i}^{q}$
over all the data points within the box $I_{1,l_{1}}\!\times\ldots\times I_{d,l_{d}}$.
\end{lem}
\begin{figure}[h]
\begin{centering}
\includegraphics[width=0.75\paperwidth]{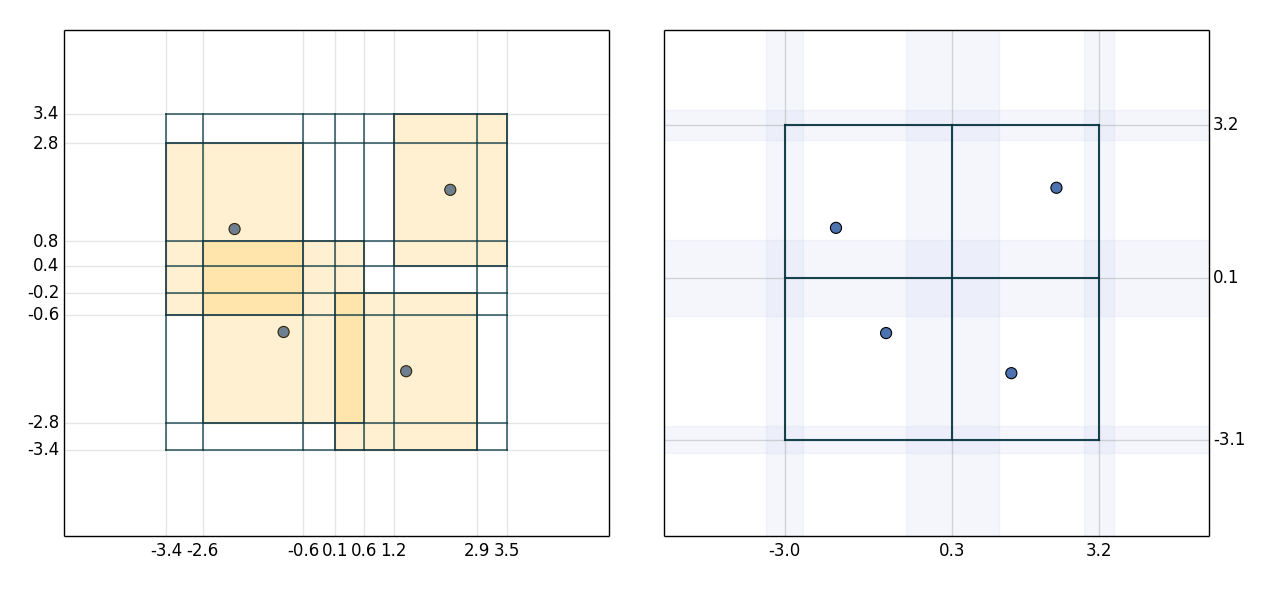}
\par\end{centering}
\caption{From bandwidths to partition (2D)\label{fig:2D_partition}}
\end{figure}

To complement the illustration of univariate partition given by Figure
\ref{fig:1D_partition}, Figure \ref{fig:2D_partition} provides a
bivariate partition example. There are four points, each at the center
of their respective rectangular kernel (in orange). On the left-hand
side, the bandwidths boundaries are used to produce the partitions
$\tilde{\mathcal{G}}_{k}$ in each dimension. One can see that most
of the resulting rectangles are empty. On the right-hand side, the
empty rectangles are removed/merged, resulting in the trimmed partitions
$\mathcal{G}_{k}$ in each dimension. Remark that this is a simple
example for which every final rectangle only contains one point.

\subsubsection{Fast multivariate sweeping algorithm\label{subsec:Fast-multivariate-sweeping}}

So far, we have shown that computing multivariate kernel smoothers
is based on the computation of the kernel sums \eqref{eq:sum_multidim},
which can be decomposed into sums of the type \eqref{eq:SpqLRd},
which can themselves be decomposed into the smaller sums \eqref{eq:sum_partition_slim}
by decomposing every kernel support of every evaluation point onto
the box partition described in the previous subsection \ref{subsec:Data-partition}.
The final task is to define an efficient algorithm to traverse all
the hyperrectangle unions $\bigcup_{(l_{1}\!,\ldots,l_{d})\in\{L_{1,j_{1}}\!,\ldots,R_{1,j_{1}}\!\}\!\times\ldots\times\!\{L_{d,j_{d}}\!,\ldots,R_{d,j_{d}}\!\}}I_{1,l_{1}}\!\times\ldots\times I_{d,l_{d}}$,
so as to compute the right-hand side sums in equation \eqref{eq:sum_partition_slim}
(Lemma \ref{lem:sum_decomposition}) in an efficient fast sum updating
way similar to the univariate updating \eqref{eq:1d_updating}. We
precompute all the sums $\mathcal{S}_{l_{1},l_{2},\ldots,l_{d}}^{\mathrm{idx}}$
with $\mathrm{idx}=\left(\mathbf{p},q,\mathbf{k}\right)\in\{0,1,2\}\times\{0,1\}^{3}\times\{1,2,\ldots,d\}^{3}$,
and use them as input material for fast multivariate sum updating.

We start with the bivariate case, summarized in Algorithm \ref{alg:fast2Dkreg},
with the help of Figures \ref{fig:2D_updating_1} and \ref{fig:2D_updating_2}.
We first provide an algorithm to compute the sums $\mathcal{T}_{1,l_{2}}^{\mathrm{idx}}:=\sum_{l_{1}=L_{1,j_{1}}}^{R_{1,j_{1}}}\mathcal{S}_{l_{1},l_{2}}^{\mathrm{idx}}$
, for every $l_{2}\in\left\{ 1,2,\ldots,m_{2}-1\right\} $ and every
index interval $\left[L_{1,j_{1}},R_{1,j_{1}}\right]$, $j_{1}\in\{1,2,\ldots,M_{1}\}$.
Starting with $j_{1}=1$, we first compute $\mathcal{T}_{1,l_{2}}^{\mathrm{idx}}=\sum_{l_{1}=L_{1,1}}^{R_{1,1}}\mathcal{S}_{l_{1},l_{2}}^{\mathrm{idx}}$
for every $l_{2}\in\left\{ 1,2,\ldots,m_{2}-1\right\} $. Then we
iteratively increment $j_{1}$ from $j_{1}=1$ to $j_{1}=M_{1}$.
After each incrementation of $j_{1}$, we update $\mathcal{T}_{1,l_{2}}^{\mathrm{idx}}$
by fast sum updating
\begin{equation}
\sum_{l_{1}=L_{1,j_{1}}}^{R_{1,j_{1}}}\mathcal{S}_{l_{1},l_{2}}^{\mathrm{idx}}=\sum_{l_{1}=L_{1,j_{1}-1}}^{R_{1,j_{1}-1}}\mathcal{S}_{l_{1},l_{2}}^{\mathrm{idx}}+\sum_{l_{1}=R_{1,j_{1}-1}+1}^{R_{1,j_{1}}}\mathcal{S}_{l_{1},l_{2}}^{\mathrm{idx}}-\sum_{l_{1}=L_{1,j_{1}-1}}^{L_{1,j_{1}}-1}\mathcal{S}_{l_{1},l_{2}}^{\mathrm{idx}}\label{eq:2d_updating_1}
\end{equation}

The second stage is to perform a fast sum updating in the second dimension,
with the sums $\mathcal{T}_{1,l_{2}}^{\mathrm{idx}}=\sum_{l_{1}=L_{1,j_{1}}}^{R_{1,j_{1}}}\mathcal{S}_{l_{1},l_{2}}^{\mathrm{idx}}$
as input material. Our goal is to compute the sums $\mathcal{T}_{2}^{\mathrm{idx}}:=\sum_{l_{2}=L_{2,j_{2}}}^{R_{2,j_{2}}}\mathcal{T}_{1,l_{2}}^{\mathrm{idx}}$
for every index interval $\left[L_{2,j_{2}},R_{2,j_{2}}\right]$,
$j_{2}\in\{1,2,\ldots,M_{2}\}$. In a similar manner, we start from
$j_{2}=1$ with the initial sum $\mathcal{T}_{2}^{\mathrm{idx}}=\sum_{l_{2}=L_{2,1}}^{R_{2,1}}\mathcal{T}_{1,l_{2}}^{\mathrm{idx}}$.
We then increment $j_{2}$ from $j_{2}=1$ to $j_{2}=M_{2}$ iteratively.
After each incrementation of $j_{2}$, we update $\mathcal{T}_{2}^{\mathrm{idx}}$
by fast sum updating:
\begin{equation}
\sum_{l_{2}=L_{2,j_{2}}}^{R_{2,j_{2}}}\mathcal{T}_{1,l_{2}}^{\mathrm{idx}}=\sum_{l_{2}=L_{2,j_{2}-1}}^{R_{2,j_{2}-1}}\mathcal{T}_{1,l_{2}}^{\mathrm{idx}}+\sum_{l_{2}=R_{2,j_{2}-1}+1}^{R_{2,j_{2}}}\mathcal{T}_{1,l_{2}}^{\mathrm{idx}}-\sum_{l_{2}=L_{2,j_{2}-1}}^{L_{2,j_{2}}-1}\mathcal{T}_{1,l_{2}}^{\mathrm{idx}}\label{eq:2d_updating_2}
\end{equation}
Using Lemma \ref{lem:sum_decomposition} (equation \eqref{eq:sum_partition_slim}),
the resulting sum $\sum_{l_{2}=L_{2,j_{2}}}^{R_{2,j_{2}}}\mathcal{T}_{1,l_{2}}^{\mathrm{idx}}=\sum_{l_{1}=L_{1,j_{1}}}^{R_{1,j_{1}}}\sum_{l_{2}=L_{2,j_{2}}}^{R_{2,j_{2}}}\mathcal{S}_{l_{1},l_{2}}^{\mathrm{idx}}$
is equal to $\mathcal{S}^{\mathrm{idx}}([z_{j}-h_{j},z_{j}+h_{j}])$,
which can be used to compute the kernel sums $\mathbf{S}_{j}$ using
equation \eqref{eq:parabolic-kernel-development-multivariate}, from
which the bivariate kernel smoothers (kernel density estimator, kernel
regression, locally linear regression) can be computed.

This ends the description of the fast sum updating algorithm in the
bivariate case. A graphical description of it is available in Appendix
\ref{subsec:Fast-bivariate-sweeping}. The reason for enforcing Condition
\ref{cond:=00005BEvaluation-grid=00005D} and Condition \ref{cond:=00005BKernel-support=00005D}
is now clear: they pave the way for the box partition described in
subsection \ref{subsec:Data-partition}, from which the iterative
fast sum updating, one dimension at a time, displayed on Figures \ref{fig:2D_updating_1}
and \ref{fig:2D_updating_2}, can cover all the multivariate bandwidths
of all the evaluation points on the evaluation grid.

Finally, the general multivariate case is a straightforward extension
of the bivariate case, and is summarized in Algorithm \ref{alg:fastdDkreg}.
\clearpage{}

\begin{algorithm2e}[H]
\DontPrintSemicolon
\SetAlgoLined
\vspace{1mm} 
 \KwIn{precomputed sums $\mathcal{S}_{l_{1},l_{2}}^{\mathrm{idx}}$}
  \vspace{1mm}
  $\mathrm{iL_{1}}$ = 1\\
  $\mathrm{iR_{1}}$ = 1\\
  $\mathcal{T}_{1,l_{2}}^{\mathrm{idx}}$ = 0\\
 \For{$j_1=1,...,M_1$}{
  \While{( $\mathrm{iR_{1}}<m_{1}$ ) and ( $\mathrm{iR_{1}}\leq R_{1,j_{1}}$ )}{
   \vspace{0.5mm}
   $\mathcal{T}_{1,l_{2}}^{\mathrm{idx}}=\mathcal{T}_{1,l_{2}}^{\mathrm{idx}}+\mathcal{S}_{\mathrm{iR_{1}},l_{2}}^{\mathrm{idx}}$, $\forall l_{2}\in\left\{ 1,2,\ldots,m_{2}-1\right\} $\\
  \vspace{0.5mm}
   $\mathrm{iR_{1}}=\mathrm{iR_{1}}+1$\\
  }
  \While{( $\mathrm{iL_{1}}<m_{1}$ ) and ( $\mathrm{iL_{1}}<L_{1,j_{1}}$ )}{
   \vspace{0.5mm}
   $\mathcal{T}_{1,l_{2}}^{\mathrm{idx}}=\mathcal{T}_{1,l_{2}}^{\mathrm{idx}}-\mathcal{S}_{\mathrm{iL_{1}},l_{2}}^{\mathrm{idx}}$, $\forall l_{2}\in\left\{ 1,2,\ldots,m_{2}-1\right\} $\\
   \vspace{0.5mm}
   $\mathrm{iL_{1}}=\mathrm{iL_{1}}+1$\\
  }
  \Comment*[l]{Here $\mathcal{T}_{1,l_{2}}^{\mathrm{idx}}=\sum_{l_{1}=L_{1,j_{1}}}^{R_{1,j_{1}}}\mathcal{S}_{l_{1},l_{2}}^{\mathrm{idx}}$, $\forall l_{2}\in\left\{ 1,2,\ldots,m_{2}-1\right\} $}
  \vspace{1mm}
  $\mathrm{iL_{2}}$ = 1\\
  $\mathrm{iR_{2}}$ = 1\\
  $\mathcal{T}_{2}^{\mathrm{idx}}$ = 0\\
   \For{$j_2=1,...,M_2$}{
  \While{( $\mathrm{iR_{2}}<m_{2}$ ) and ( $\mathrm{iR_{2}}\leq R_{2,j_{2}}$ )}{
   \vspace{0.5mm}
   $\mathcal{T}_{2}^{\mathrm{idx}}=\mathcal{T}_{2}^{\mathrm{idx}}+\mathcal{T}_{1,\mathrm{iR_{2}}}^{\mathrm{idx}}$\\
   \vspace{0.5mm}
   $\mathrm{iR_{2}}=\mathrm{iR_{2}}+1$\\
  }
  \While{( $\mathrm{iL_{2}}<m_{2}$ ) and ( $\mathrm{iL_{2}}<L_{2,j_{2}}$ )}{
   \vspace{0.5mm}
   $\mathcal{T}_{2}^{\mathrm{idx}}=\mathcal{T}_{2}^{\mathrm{idx}}-\mathcal{T}_{1,\mathrm{iL_{2}}}^{\mathrm{idx}}$\\
   \vspace{0.5mm}
   $\mathrm{iL_{2}}=\mathrm{iL_{2}}+1$\\
  }
  \Comment*[l]{Here $\mathcal{T}_{2}^{\mathrm{idx}}=\sum_{l_{1}=L_{1,j_{1}}}^{R_{1,j_{1}}}\sum_{l_{2}=L_{2,j_{2}}}^{R_{2,j_{2}}}\mathcal{S}_{l_{1},l_{2}}^{\mathrm{idx}}$}
\vspace{0.5mm}
\Comment*[l]{$=\mathcal{S}_{\mathbf{k}}^{\mathbf{p},q}([z_{j}-h_{j},z_{j}+h_{j}])$ from equation \eqref{eq:sum_partition_slim}}
  \vspace{1mm}
  Compute $\mathbf{S}_{j}$ using $\mathcal{T}_{2}^{\mathrm{idx}}$ and equation \eqref{eq:parabolic-kernel-development-multivariate}\\
  \vspace{0.5mm}
  Compute multivariate kernel smoothers using $\mathbf{S}_{j}$  
 }
 }
\vspace{1mm}
 \KwOut{Bivariate kernel smoothers}
\vspace{1mm}
 \caption{Fast bivariate kernel smoothing\label{alg:fast2Dkreg}}
 \end{algorithm2e}

\clearpage{}

\begin{algorithm2e}[H]
\DontPrintSemicolon
\SetAlgoLined
\vspace{1mm} 
 \KwIn{precomputed sums $\mathcal{S}_{l_{1},l_{2},\ldots,l_{d}}^{\mathrm{idx}}$}
  \vspace{0.5mm}
  $\mathrm{iL_{1}}$ = 1, $\mathrm{iR_{1}}$ = 1, $\mathcal{T}_{1,l_{2},l_{3},\ldots,l_{d}}^{\mathrm{idx}}$ = 0\\
 \For{$j_1=1,...,M_1$}{
  \While{( $\mathrm{iR_{1}}<m_{1}$ ) and ( $\mathrm{iR_{1}}\leq R_{1,j_{1}}$ )}{
   \vspace{0.5mm}
   $\mathcal{T}_{1,l_{2},l_{3},\ldots,l_{d}}^{\mathrm{idx}}\!=\mathcal{T}_{1,l_{2},l_{3},\ldots,l_{d}}^{\mathrm{idx}}\!+\mathcal{S}_{\mathrm{iR_{1}},l_{2},l_{3},\ldots,l_{d}}^{\mathrm{idx}}$,${\scriptstyle \forall l_{k}\in\left\{ 1,2,\ldots,m_{k}-1\right\} }$,${\scriptstyle k\in\{2,3,\ldots,d\}}$\\
  \vspace{0.5mm}
   $\mathrm{iR_{1}}=\mathrm{iR_{1}}+1$\\
  }
  \While{( $\mathrm{iL_{1}}<m_{1}$ ) and ( $\mathrm{iL_{1}}<L_{1,j_{1}}$ )}{
   \vspace{0.5mm}
   $\mathcal{T}_{1,l_{2},l_{3},\ldots,l_{d}}^{\mathrm{idx}}\!=\mathcal{T}_{1,l_{2},l_{3},\ldots,l_{d}}^{\mathrm{idx}}\!-\mathcal{S}_{\mathrm{iL_{1}},l_{2},l_{3},\ldots,l_{d}}^{\mathrm{idx}}$,${\scriptstyle \forall l_{k}\in\left\{ 1,2,\ldots,m_{k}-1\right\} }$,${\scriptstyle k\in\{2,3,\ldots,d\}}$\\
   \vspace{0.5mm}
   $\mathrm{iL_{1}}=\mathrm{iL_{1}}+1$\\
  }
  \vspace{-1mm}
  \Comment*[l]{Here $\mathcal{T}_{1,l_{2},l_{3},\ldots,l_{d}}^{\mathrm{idx}}\!=\sum_{l_{1}=L_{1,j_{1}}}^{R_{1,j_{1}}}\!\mathcal{S}_{l_{1},l_{2},\ldots,l_{d}}^{\mathrm{idx}}$,${\scriptstyle \forall l_{k}\in\left\{ 1,2,\ldots,m_{k}-1\right\} }$,${\scriptstyle k\in\{2,3,\ldots,d\}}$}
  \vspace{1.0mm}
  $\mathrm{iL_{2}}$ = 1, $\mathrm{iR_{2}}$ = 1, $\mathcal{T}_{2,l_{3},\ldots,l_{d}}^{\mathrm{idx}}$ = 0\\
   \For{$j_2=1,...,M_2$}{
  \While{( $\mathrm{iR_{2}}<m_{2}$ ) and ( $\mathrm{iR_{2}}\leq R_{2,j_{2}}$ )}{
   \vspace{0.5mm}
   $\mathcal{T}_{2,l_{3},\ldots,l_{d}}^{\mathrm{idx}}\!=\mathcal{T}_{2,l_{3},\ldots,l_{d}}^{\mathrm{idx}}\!+\mathcal{T}_{1,\mathrm{iR_{2}},l_{3},\ldots,l_{d}}^{\mathrm{idx}}$, ${\scriptstyle \forall l_{k}\in\left\{ 1,2,\ldots,m_{k}-1\right\} }$, ${\scriptstyle k\in\{3,\ldots,d\}}$\\
   \vspace{0.5mm}
   $\mathrm{iR_{2}}=\mathrm{iR_{2}}+1$\\
  }
  \While{( $\mathrm{iL_{2}}<m_{2}$ ) and ( $\mathrm{iL_{2}}<L_{2,j_{2}}$ )}{
   \vspace{0.5mm}
   $\mathcal{T}_{2,l_{3},\ldots,l_{d}}^{\mathrm{idx}}\!=\mathcal{T}_{2,l_{3},\ldots,l_{d}}^{\mathrm{idx}}\!-\mathcal{T}_{1,\mathrm{iL_{2}},l_{3},\ldots,l_{d}}^{\mathrm{idx}}$, ${\scriptstyle \forall l_{k}\in\left\{ 1,2,\ldots,m_{k}-1\right\} }$, ${\scriptstyle k\in\{3,\ldots,d\}}$\\
   \vspace{0.5mm}
   $\mathrm{iL_{2}}=\mathrm{iL_{2}}+1$\\
  }
  \vspace{-1.0mm}
  \Comment*[l]{\!\!$\mathcal{T}_{2,l_{3},\ldots,l_{d}}^{\mathrm{idx}}\!=\sum_{l_{1}=L_{1,j_{1}}}^{R_{1,j_{1}}}\!\sum_{l_{2}=L_{2,j_{2}}}^{R_{2,j_{2}}}\!\!\mathcal{S}_{l_{1},l_{2},\ldots,l_{d}}^{\mathrm{idx}}$,${\scriptstyle \forall l_{k}\in\left\{ 1,\ldots,m_{k}-1\right\} }$,${\scriptstyle k\in\{3,\ldots d\}}$}
  \vspace{-1.5mm}
  \vdots
  $\mathrm{iL_{d}}$ = 1, $\mathrm{iR_{d}}$ = 1, $\mathcal{T}_{d}$ = 0\\
   \For{$j_d=1,...,M_d$}{
  \While{( $\mathrm{iR_{d}}<m_{d}$ ) and ( $\mathrm{iR_{d}}\leq R_{d,j_{d}}$ )}{
   \vspace{0.5mm}
   $\mathcal{T}_{d}^{\mathrm{idx}}=\mathcal{T}_{d}^{\mathrm{idx}}+\mathcal{T}_{d-1,\mathrm{iR_{d}}}^{\mathrm{idx}}$\\
   \vspace{0.5mm}
   $\mathrm{iR_{d}}=\mathrm{iR_{d}}+1$\\
  }
  \While{( $\mathrm{iL_{d}}<m_{d}$ ) and ( $\mathrm{iL_{d}}<L_{d,j_{d}}$ )}{
   \vspace{0.5mm}
   $\mathcal{T}_{d}^{\mathrm{idx}}=\mathcal{T}_{d}^{\mathrm{idx}}-\mathcal{T}_{d-1,\mathrm{iL_{d}}}^{\mathrm{idx}}$\\
   \vspace{0.5mm}
   $\mathrm{iL_{d}}=\mathrm{iL_{d}}+1$\\
  }
  \vspace{-1.0mm}
  \Comment*[l]{Here $\mathcal{T}_{d}^{\mathrm{idx}}=\sum_{l_{1}=L_{1,j_{1}}}^{R_{1,j_{1}}}\sum_{l_{2}=L_{2,j_{2}}}^{R_{2,j_{2}}}\cdots\sum_{l_{d}=L_{d,j_{d}}}^{R_{d,j_{d}}}\mathcal{S}_{l_{1},l_{2},\ldots,l_{d}}^{\mathrm{idx}}$}
\vspace{0.5mm}
\Comment*[l]{$=\mathcal{S}_{\mathbf{k}}^{\mathbf{p},q}([z_{j}-h_{j},z_{j}+h_{j}])$ from equation \eqref{eq:sum_partition_slim}}
  \vspace{1mm}
  Compute $\mathbf{S}_{j}$ using $\mathcal{T}_{d}^{\mathrm{idx}}$ and equation \eqref{eq:parabolic-kernel-development-multivariate}\\
  \vspace{0.5mm}
  Compute multivariate kernel smoothers using $\mathbf{S}_{j}$  
 }\vspace{-1mm}
 }\vspace{-1mm}
 }
\vspace{0.0mm}
 \KwOut{Multivariate kernel smoothers}
\vspace{1mm}
 \caption{Fast multivariate kernel smoothing\label{alg:fastdDkreg}}
 \end{algorithm2e}

\subsection{Complexity\label{subsec:Complexity}}

\subsubsection{Computational complexity}

One can verify that the number of operations in the multivariate fast
sum updating algorithm \ref{alg:fastdDkreg} is proportional to the
number of evaluation points $M=M_{1}\times M_{2}\times\ldots\times M_{d}$.
Indeed, recall from subsection \ref{subsec:Data-partition} that in
each dimension $k\in\left\{ 1,2,\ldots,d\right\} $, $m_{k}-1$ is
the number of intervals in the $k$-th dimension of the data partition,
with $2\leq m_{k}\leq2M_{k}$. The first two \textsf{\textit{while}}
loops over $\mathrm{iR_{1}}$ and $\mathrm{iL_{1}}$ in Algorithm
\ref{alg:fastdDkreg} generate $2(m_{1}-1)$ updates of the sums $\mathcal{T}_{1,l_{2},l_{3},\ldots,l_{d}}^{\mathrm{idx}}$
of size $(m_{2}-1)\times\ldots\times(m_{d}-1)$, for a total of $\mathcal{O}(M)$
operations. Then, the two subsequent\textsf{\textit{ while}} loops
over $\mathrm{iR_{2}}$ and $\mathrm{iL_{2}}$ generate $M_{1}\times2(m_{2}-1)$
updates of the sums $\mathcal{T}_{2,l_{3},\ldots,l_{d}}^{\mathrm{idx}}$
of size $(m_{3}-1)\times\ldots\times(m_{d}-1)$, for a total of $\mathcal{O}(M)$
operations. The final \textsf{\textit{while}} loops over $\mathrm{iR_{d}}$
and $\mathrm{iL_{d}}$ generate $M_{1}\times\ldots\times M_{d-1}\times2(m_{d}-1)=\mathcal{O}(M)$
updates of the sum $\mathcal{T}_{d}^{\mathrm{idx}}$ of size $1$.
The computational complexity of Algorithm \ref{alg:fastdDkreg} is
therefore $\mathcal{O}(M)$. 

In addition to this cost, Algorithm \ref{alg:fastdDkreg} requires
the construction of the partition $\mathcal{G}_{k}$ and of the threshold
indices $L_{k,j_{k}}\in\{1,2,\ldots,m_{k}-1\}$ and $R_{k,j_{k}}\in\{1,2,\ldots,m_{k}-1\}$
(recall Lemma \ref{lem:sum_decomposition}), which costs $\mathcal{O}(M)$
operations or $\mathcal{O}(M\log M)$ is the evaluation points are
not sorted. The precomputation of the sums $\mathcal{S}_{l_{1},l_{2},\ldots,l_{d}}^{\mathrm{idx}}$
costs $\mathcal{O}(N)$ operations once the input sample $\left(x_{1,i},x_{2,i},\ldots,x_{d,i}\right)$,
$i\in\{1,2,\ldots,N\}$ has been sorted in each dimension independently,
at a cost of $\mathcal{O}(N\log N)$ operations. The total computational
complexity of the multivariate fast sum updating algorithm described
in this section is therefore $\mathcal{O}(M\log M+N\log N)$, which
is a considerable improvement over the $\mathcal{O}(M\times N)$ complexity
of the naive approach.

\subsubsection{Memory complexity}

The memory comsumption of Algorithm \ref{alg:fastdDkreg} stems from
the simultaneous storage of the sums $\mathcal{S}_{l_{1},l_{2},\ldots,l_{d}}^{\mathrm{idx}}$,
$\mathcal{T}_{1,l_{2},l_{3},\ldots,l_{d}}^{\mathrm{idx}}$, $\mathcal{T}_{2,l_{3},\ldots,l_{d}}^{\mathrm{idx}}$,
$\ldots$, $\mathcal{T}_{d}^{\mathrm{idx}}$ for every $l_{k}\!\in\!\left\{ 1,2,\!\ldots\!,m_{k}-1\right\} $,
$k\in\{2,3,\ldots,d\}$ and $\mathrm{idx}=\left(\mathbf{p},q,\mathbf{k}\right)\in\{0,1,2\}\times\{0,1\}^{3}\times\{1,2,\ldots,d\}^{3}$,
resulting in a memory complexity of $\mathcal{O}(M)$.

\subsubsection{Dependence in $d$}

Finally, we look at the dependence in the dimension $d$ of the constant
in the computational and memory complexities of the algorithm. In
the worst case, $m_{k}$ is equal to its upper bound $2M_{k}$ in
every dimension $k\in\left\{ 1,2,\ldots,d\right\} $. In such a case,
Algorithm \ref{alg:fastdDkreg} generates $\mathcal{O}(2^{d}M)$ operations
for a single index calculation resulting in a global cost in $\mathcal{O}(d^{3}2^{d}M)$,
where $d^{3}$ comes from the dimension of the multi-index $\mathrm{idx}=\left(\mathbf{p},q,\mathbf{k}\right)\in\{0,1,2\}\times\{0,1\}^{3}\times\{1,2,\ldots,d\}^{3}$
as well as the fact that solving the regression system \eqref{eq:localreg1-d-explicit}
costs $\mathcal{O}(d^{3})$ operations for each evaluation point $z_{j}$,
$j\in\{1,2,\ldots,M\}$. In practice, the constant $2^{d}$ can be
greatly reduced depending on the size of the slimmed down partition
$\{\mathcal{G}_{k}\}_{k=1,\ldots,d}$ compared to the initial partition
$\{\tilde{\mathcal{G}}_{k}\}_{k=1,\ldots,d}$. Similarly, the worst
case memory storage needed for the sums $\mathcal{S}_{l_{1},l_{2},\ldots,l_{d}}^{\mathrm{idx}}$
and the terms $\mathcal{T}_{1,l_{2},l_{3},\ldots,l_{d}}^{\mathrm{idx}}$,
$\mathcal{T}_{2,l_{3},\ldots,l_{d}}^{\mathrm{idx}}$, $\ldots$, $\mathcal{T}_{d}^{\mathrm{idx}}$
is $\mathcal{O}(d^{3}2^{d}M)$ where again the constant $2^{d}$ can
be greatly reduced in practice. The sorting of the input points and
evaluation points in each dimension and the precomputation of sums
and indices generate altogether $\mathcal{O}(dM\log M+dN\log N)$
operations. The total computational complexity is therefore $\mathcal{O}(dM(\log M+d^{2}2^{d})+dN\log N)$
in the worst case, with the term $2^{d}$ possibly smaller in practice. 

By contrast, the naive approach requires $\mathcal{O}(d^{2}M\times(N+d))$
operations: for each $j\in\{1,2,\ldots,M\}$, one needs $\mathcal{O}(dN)$
for computing $K_{d,h}(x_{i}-z_{j})$ for each $i\in\{1,2,\ldots,N\}$
(equation \eqref{eq:kernel_mean}), $\mathcal{O}(d^{2}N)$ for computing
all the sums in \eqref{eq:localreg1-d-explicit} and $\mathcal{O}(d^{3})$
for solving the system \eqref{eq:localreg1-d-explicit}. This shows
that the multivariate fast sum updating is faster than the naive approach
whenever $d2^{d}\ll N$, and still likely to be faster beyond this
case as the $2^{d}$ constant only occurs in the unlikely worst case
scenario for which $m_{k}=2M_{k}$ in each dimension $k\in\{1,2,\ldots,d\}$. 

\section{Evaluation grid and adaptive bandwidth\label{sec:grid-bandwidth}}

This section suggests some suitable choices of evaluation grid and
adaptive bandwidth compatible with the two conditions \ref{cond:=00005BEvaluation-grid=00005D}
and \ref{cond:=00005BKernel-support=00005D}, so as to ensure a wide
applicability of the fast kernel smoothers described in this paper.

\subsection{Shape of evaluation grid\label{subsec:grid-shape}}

\begin{figure}[h]
\begin{centering}
\includegraphics[width=0.7\paperwidth]{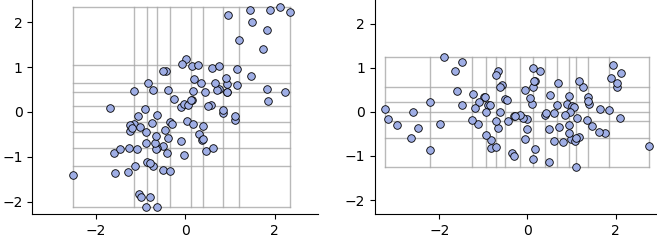}
\par\end{centering}
\caption{Evaluation grid: rotation\label{fig:evaluation grid}}
\end{figure}

As explained in \ref{subsec:Multivariate-kernel-smoothers}, kernel
smoothing estimates can generally be improved by prerotating the input
dataset into a better basis. Rotating the dataset before performing
kernel density estimations has been advocated in \citet{Wand94} and
\citet{Scott05} for example. Condition \ref{cond:=00005BEvaluation-grid=00005D}
adds another motivation for rotating the dataset. Indeed, when the
natural evaluation sample is not a grid, for example when the evaluation
points $z_{1}$, $z_{2}$, $\ldots$, $z_{M}$ are equal to the input
points $x_{1}$, $x_{2}$, $\ldots$, $x_{N}$, one needs to build
a suitable intermediate evaluation grid to properly cover the input
sample. The left-side of Figure \ref{fig:evaluation grid} illustrates
on a bivariate example with $N=100$ input points the potential problem
of rectilinear evaluation grids when the dimensions of the input dataset
are dependent: some evaluation points can be left away from the dataset.
As some evaluation points are located in empty areas, the effective
number of evaluation points is decreased. A rotation of the input
dataset can mitigate or eliminate this problem, as shown on the right-side
of Figure \ref{fig:evaluation grid}.

To construct the rotation, several techniques can be used. One possible
choice is to rotate the dataset onto its principal components (as
shown on Figure \ref{fig:evaluation grid}). To define the evaluation
grid $\left\{ (z_{1,j_{1}},z_{2,j_{2}},\ldots,z_{d,j_{d}}),\,j_{k}\in\{1,2,\ldots,M_{k}\},\,k\in\{1,2,\ldots,d\}\right\} $,
one can first set each $M_{k}$ to $M^{\frac{1}{d}}$ and define 
\[
z_{k,j_{k}}=x_{k,\mathrm{round}\left(1+(N-1)\times\frac{j_{k}-1}{M_{k}-1}\right)}
\]
where $\mathrm{round}(u)$ is the closest integer to $u\in\mathbb{R}$
and the input set $x_{1}$, $x_{2}$, $\ldots$, $x_{N}$ is sorted
in increasing order. Such a grid is illustrated on the left-side of
Figure \ref{fig:evaluation grid} (original dataset without rotation).
One alternative choice for $M_{k}$ is to set it proportional to the
$k$-th singular value associated with the $k$-th principal component.
In addition to improving the coverage of the input dataset by the
evaluation grid (more evaluation points along the more variable input
dimensions), this choice of $M_{k}$ can reduce the dimension of the
problem whenever some $M_{k}$ are set to one due to a small singular
value. This choice of $M_{k}$ is illustrated on the right-side of
Figure \ref{fig:evaluation grid} (after the rotation of the input
dataset), and is the one we use in the numerical section \ref{sec:Numerical}.
Once the kernel density estimates have been obtained on the intermediate
evaluation grid using the fast algorithm described in Section \ref{sec:Fast-sum-updating},
one can interpolate the estimates from the grid to the evaluation
points of interest by simple multilinear interpolation. Alternatively,
one can interpolate by Inverse Distance Weighting (\citet{Shepherd68}).
When the weights are chosen as kernels from Appendix \ref{sec:fast-sum-kernels},
this interpolation bears some similarity with kernel density estimation,
and can benefit from the fast sum updating algorithm described in
Section \ref{sec:Fast-sum-updating}. 

\subsection{Fast adaptive bandwidth\label{subsec:knn-bandwidth}}

The kernel smoothers \eqref{eq:localdens}, \eqref{eq:localreg0}
and \eqref{eq:localreg1} can be defined with a fixed bandwidth $h$,
or with an adaptive bandwidth which varies with either the input points
or the evaluation points. When the input design is random as on Figure
\ref{fig:evaluation grid}, some areas might be sparse while others
will be dense. In such cases, the benefit of adaptive bandwidth is
that one can maintain a uniform quality of density estimates by using
a larger bandwidth in sparse areas and a smaller bandwidth in dense
areas. There exists two main ways to define adaptive bandwidths: balloon
bandwidths $h=h_{j}$ which vary with the evaluation point $j\in\{1,2,\ldots,M\}$,
and sample point bandwidths $h=h_{i}$ which vary with the input point
$i\in\{1,2,\ldots,N\}$, see \citet{Terrell92} or \citet{Scott05}.
While many univariate kernels in Table \ref{tab:fast-sum-kernels}
are compatible with both balloon bandwidths and sample point bandwidths
(see Appendix \ref{sec:fast-sum-kernels}), the data partition from
subsection \ref{subsec:Data-partition}, which is required in the
multivariate case, has been tailored for the balloon formulation (Condition
\ref{cond:=00005BKernel-support=00005D}), which is the one we adopt
in this paper.

For the construction of the adaptive bandwidth, we adopt the $K$-nearest
neighbor bandwidth suggested in \citet{Loftsgaarden65}, as it was
shown in \citet{Terrell92} to perform well in multivariate settings.
In addition, such a choice of bandwidth ensures that the bandwidth
boundaries $\{z_{k,j_{k}}-h_{k,j_{k}}\}_{j_{k}=1,\ldots,M_{k}}$ and
$\{z_{k,j_{k}}+h_{k,j_{k}}\}_{j_{k}=1,\ldots,M_{k}}$, $k=\{1,2,\ldots,d\}$
remain in increasing order, which was implicitly assumed in Algorithms
\ref{alg:fast2Dkreg} and \ref{alg:fastdDkreg} for simplicity (one
can easily adjust the loops to decrement instead of increment the
grid indices $\mathrm{iL_{k}}$ and $\mathrm{iR_{k}}$ whenever the
bandwidth boundaries are not in increasing order).

We now describe how to build these bandwidths in a fast $\mathcal{O}(M+N)$
from sorted datasets in the univariate case ($\mathcal{O}(M\log M+N\log N)$
if the datasets need to be sorted beforehand), and then discuss the
extension to the multivariate case.

Let $x_{1}\leq x_{2}\leq\ldots\leq x_{N}$ be a sorted set of $N$
sample points, and $z_{1}\leq z_{2}\leq\ldots\leq z_{M}$ be a sorted
set of $M$ evaluation points. Algorithm \ref{alg:fast1Dknn} describes
an efficient algorithm to build $M$ adaptive bandwidths $h_{j}$
centered around the points $z_{j}$, $j=1,\ldots,M$, such that each
bandwidth $\left[z_{j}-h_{j},z_{j}+h_{j}\right]$ contains exactly
$K$ sample points.\vspace{1mm}

\begin{algorithm2e}[H]
\DontPrintSemicolon
\SetAlgoLined
\vspace{1mm} 
 \KwIn{\\
  X: sorted vector of N real points $X[1]\leq \ldots\leq X[N]$\\
  Z: sorted vector of M evaluation points $Z[1]\leq \ldots\leq Z[M]$\\
  K: number of points that each bandwidth should include $(1\leq K\leq N)$ }  
  \vspace{1mm}
\Comment*[l]{The indices iL and iR define a subset $X[iL], X[iL+1],..., X[iR]$ of K points}  
 iL = $1$ \Comment*[l]{Left index}   
 iR = $K$ \Comment*[l]{Right index} 
 cM = $(X[iL]+X[iR+1])/2$ \Comment*[l]{Middle cut} 
 dmax = $X[N]-X[1]$\Comment*[l]{Maximum distance between 2 sample points}
\vspace{1mm}
 \For{$i=1,...,M$}{
  \While{(iR+1$<$N) and (Z[i]$>$cM)}{
   iL = iL+1\\
   iR = iR+1\\
   cM = $0.5*(X[iL]+X[iR+1])$\\
  }
  Hmin = $\max\left\{\ Z[i]-X[iL]\ ,\ X[iR]-Z[i]\ \right\}$\\
  Hmax = $\min\left\{\ (Z[i]-X[iL-1])\times\mathbbm{1}\{iL>1\} + \mathrm{dmax}\times\mathbbm{1}\{iL=1\}\ ,\right.$\\
  \hspace{6.6em}$\left. (X[iR+1]-Z[i])\times\mathbbm{1}\{iR<N\} + \mathrm{dmax}\times\mathbbm{1}\{iR=N\}\ \right\}$\\
  H[i] = (Hmin+Hmax)/2\\
 }
return H\\ 
\vspace{1mm}
 \KwOut{\\H: for each point i, the interval $[Z[i]-H[i] , Z[i]+H[i]]$ contains exactly K points}
\vspace{1mm}
 \caption{Fast univariate $K$-Nearest Neighbors bandwidth\label{alg:fast1Dknn}}
 \end{algorithm2e}

\vspace{1mm}
Define $i_{L}\in[1,\ldots,N-K+1]$ and $i_{R}=i_{L}+K-1$. The subset
$x_{i_{L}},x_{i_{L}+1},\ldots,x_{i_{R}}$ contains exactly $K$ points.
The idea of the algorithm is to enumerate all such possible index
ranges $\left[i_{L},i_{R}\right]$ from left ($i_{L}=1$, $i_{R}=K$)
to right ($i_{L}=N-K+1$, $i_{R}=N$), and to match each evaluation
point $z_{j}$, $j\in\{1,2,\ldots,N\}$ with its corresponding $K$-nearest-neighbors
subsample $x_{i_{L}},x_{i_{L}+1},\ldots,x_{i_{R}}$. 

Matching each index $j$ to its corresponding $\left[i_{L},i_{R}\right]$
range is simple. When $i_{L}=1$, all the points $z_{j}$ such that
$z_{j}\leq(x_{i_{L}}+x_{i_{R}+1})/2$ are such that the subsample
$x_{i_{L}},x_{i_{L}+1},\ldots,x_{i_{R}}$ corresponds to their $K$
nearest neighbors. Indeed, any point greater than $(x_{i_{L}}+x_{i_{R}+1})/2$
is closer to $x_{i_{R}+1}$ than to $x_{i_{L}}$, and therefore its
$K$ nearest neighbors are not $x_{i_{L}},x_{i_{L}+1},\ldots,x_{i_{R}}$.

Once all such $z_{j}$ are matched to the current $\left[i_{L},i_{R}\right]$
range, $i_{L}$ and $i_{R}$ are incremented until $(x_{i_{L}}+x_{i_{R}+1})/2$
is greater than the next evaluation point $z_{j}$ to assign. The
same procedure is then repeated until all the points are assigned
to their $K$ nearest neighbors.

Finally, once each point $z_{j}$ is assigned to its $K$ nearest
neighbors $x_{i_{L}},x_{i_{L}+1},\ldots,x_{i_{R}}$, one still needs
to choose the bandwidth $h_{j}$ such that $\left[z_{j}-h_{j},z_{j}+h_{j}\right]$
contains these $K$ nearest neighbors. Such a bandwidth $h_{j}$ exists
but is not unique. We choose to set $h_{j}$ to the average between
the smallest possible $h_{j}$ (equal to $\max\left\{ z_{j}-x_{i_{L}},x_{i_{R}}-z_{j}\right\} $)
and the largest possible $h_{j}$ (equal to $\min\left\{ z_{j}-x_{i_{L}-1},x_{iR+1}-z_{j}\right\} $
when $i_{L}-1\geq1$ and $i_{R}+1\leq N$).

\begin{figure}[h]
\begin{centering}
\includegraphics[width=0.75\paperwidth]{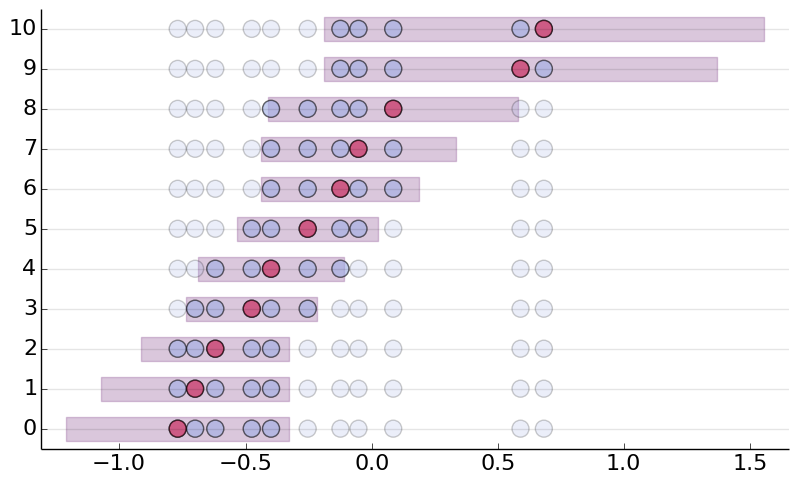}
\par\end{centering}
\caption{1D nearest neighbour bandwidth ($N=11$, $K=5$)\label{fig:1D_bandwidths}}

\end{figure}

Figure \ref{fig:1D_bandwidths} illustrates the resulting bandwidths
on a random sample of $11$ points, where the evaluation points and
the sample points are set to be the same for simplicity. Each row
repeats the whole sample, and shows the bandwidth centered around
one of the $11$ points, and containing $k=5$ points.

The computational complexity of Algorithm \ref{alg:fast1Dknn} is
a fast $\mathcal{O}(M+N)$ if the set of input points $x_{i}$ $i\in\{1,2,\ldots,N\}$
and the set of evaluation points $z_{j}$, $j\in\{1,2,\ldots,M\}$
are already sorted, and $\mathcal{O}(M\log(M)+N\log(N))$ otherwise.

In the multivariate case, given Condition \ref{cond:=00005BKernel-support=00005D},
what can be done is to compute approximate multivariate $K$-nearest-neighbors
bandwidths by performing Algorithm \ref{alg:fast1Dknn} dimension
per dimension. Let $p=K/N=p_{1}\times p_{2}\times\ldots\times p_{d}$
be the proportion of input points to include within each multivariate
bandwidth. In practice, we set $p_{k}$ to be inversely proportional
to the $k$-th singular value associated with the $k$-th axis (projected
onto $[0,1]$ if it ends up outside this probability range) and run
Algorithm \ref{alg:fast1Dknn} with $K_{k}=p_{k}\times N$ in each
dimension $k\in\{1,2,\ldots,d\}$ independently, which should ensure
each multivariate bandwidth contains approximately $K$ input points,
provided the rotation onto the principal components has been performed
beforehand (subsection \ref{subsec:grid-shape}).

\section{Numerical tests\label{sec:Numerical}}

In this section, we test the fast kernel summation algorithm introduced
in Section \ref{sec:Fast-sum-updating} and compare it to naive summation
in terms of speed and accuracy. We consider a sample of $N$ input
points, choose the number of evaluation points $M$ approximately
equal to $N$, and build the evaluation grid and the bandwidths as
described in Section \ref{sec:grid-bandwidth}. 

From subsection \ref{subsec:Complexity}, we expect a runtime proportional
to $N\log(N)$. We are going to verify this result numerically. Then,
we are going to compare the estimates obtained by fast kernel summation
to those obtain by naive summation. As discussed in subsection \ref{subsec:Numerical-stability},
we expect small differences coming from the rounding of floats, which
can be reduced or removed altogether by the use of stable summation
algorithm. As a simple illustration, we measure the accuracy improvement
provided by the simple M\o ller-Kahan algorithm (\citet{Moller65},
\citet{Linnainmaa74}, \citet{Ozawa83}, see Appendix \ref{sec:Stable-fast-sum}).
Beyond this simple stability improvement , one can instead use exact
summation algorithms to remove any float rounding errors while maintaining
the $\mathcal{O}(N\log(N))$ complexity (cf. subsection \ref{subsec:Numerical-stability}).

The input sample can be chosen arbitrarily as it does not affect the
speed or accuracy of the two algorithms. We therefore simply choose
to simulate $N$ points from a $d$-dimensional Gaussian random variable
$X\sim\mathbb{N}(0,0.6\mathbf{1_{d}})$. In addition to the input
sample $x_{1}$, $x_{2}$, $\ldots$, $x_{N}$ , we need an output
sample $y_{1}$, $y_{2}$, $\ldots$, $y_{N}$, in order to test the
locally linear regression. Similarly to the input sample, the output
sample can be chosen arbitrarily. We choose to define the output as
\begin{eqnarray*}
Y & = & f(X)+W\\
f(x) & = & \sum_{i=1}^{d}x_{i}+\exp\left(-16\left(\sum_{i=1}^{d}x_{i}\right)^{2}\right)
\end{eqnarray*}
where the univariate Gaussian noise $W\sim\mathbb{N}(0,0.7)$ is independent
of $X$. 

The various tables in this section report the following values: 
\begin{description}
\item [{Fast~kernel~time}] stands for the computational time in seconds
taken by the fast kernel summation algorithm; 
\item [{Naive~time}] stands for the computational time in seconds of the
naive version; 
\item [{Accur~Worst}] stands for the maximum relative error of the fast
sum algorithm on the whole grid. For each evaluation point, this relative
error is computed as $|E_{\mathrm{fast}}-E_{\mathrm{naive}}|/|E_{\mathrm{naive}}|$
where $E_{\mathrm{fast}}$ and $E_{\mathrm{naive}}$ are the estimates
obtained by the fast sum updating algorithm and the naive summation
algorithm, respectively; 
\item [{Accur~Worst~Stab}] stands for the maximum relative error of the
fast sum algorithm with M\o ller-Kahan stabilization on the whole
grid; 
\item [{Accur~Aver}] stands for the average relative error on the grid. 
\item [{Accur~Aver~Stab}] stands for the average relative error of the
fast sum algorithm with stabilization on the whole grid. 
\end{description}
We perform the tests on an Intel\textregistered{} Xeon\textregistered{} CPU E5-2680 v4 @ 2.40GHz (Broadwell)\footnote{https://ark.intel.com/products/91754/Intel-Xeon-Processor-E5-2680-v4-35M-Cache-2\_40-GHz}.
The code was written in C++ and is available in the StOpt\footnote{https://gitlab.com/stochastic-control/StOpt}
library (\citet{gevret2016stochastic}). Subsection \ref{subsec:kde-test}
focuses on kernel density estimation, while subsection \ref{subsec:klin-test}
considers locally linear regression. 

\subsection{Fast kernel density estimation\label{subsec:kde-test}}

This subsection focuses on kernel density estimation (equation \eqref{eq:localdens-d}).
We implement and compare the fast kernel summation and the naive summation
algorithms for different sample sizes $N$. Recalling from subsection
\ref{subsec:knn-bandwidth} that our adaptive bandwidths are defined
by the proportion $p$ of neighboring sample points to include in
each evaluation bandwidth, we test the two proportions $p=15\%$ and
$p=25\%$. 

\subsubsection{Univariate case}

We first consider the univariate case. Tables \ref{tab:1DKDE0.15}
and \ref{tab:1DKDE0.25} summarize the results obtained by Algorithm
\ref{alg:fast1Dkreg} with the two different bandwidths. The results
are very good even without stabilization, and the use of the M\o ller-Kahan
summation algorithm improves the accuracy by two digits for the same
computational cost. As expected the computational time of the fast
summation algorithm is far better than the one obtained by naive summation
(less than half a second versus more than three hours for $1.28$
million points for example).

\begin{table}[H]
\begin{centering}
\begin{tabular}{llllllll}
\toprule 
$\!\!\!$Nb particles & 20,000 & 40,000 & 80,000 & 160,000 & 320,000 & 640,000 & 1,280,000$\!\!\!$\tabularnewline
\midrule
$\!\!\!$Fast kernel time & 0.01 & 0.01 & 0.02 & 0.04 & 0.10 & 0.20 & 0.43\tabularnewline
$\!\!\!$Naive time  & 2.90 & 12 & 47 & 190 & 750 & 3,000 & 12,000\tabularnewline
\midrule 
$\!\!\!$Accur Worst & 1.7\,${\scriptstyle \mathsf{E}}$-09 & 1.1\,${\scriptstyle \mathsf{E}}$-09 & 8.2\,${\scriptstyle \mathsf{E}}$-09 & 1.1\,${\scriptstyle \mathsf{E}}$-07 & 3.2\,${\scriptstyle \mathsf{E}}$-07 & 1.7\,${\scriptstyle \mathsf{E}}$-06 & 5.0\,${\scriptstyle \mathsf{E}}$-07$\!\!\!$\tabularnewline
$\!\!\!$Accur Worst Stab  & 4.8\,${\scriptstyle \mathsf{E}}$-12 & 5.1\,${\scriptstyle \mathsf{E}}$-13 & 8.3\,${\scriptstyle \mathsf{E}}$-12 & 6.9\,${\scriptstyle \mathsf{E}}$-12 & 1.5\,${\scriptstyle \mathsf{E}}$-11 & 3.9\,${\scriptstyle \mathsf{E}}$-10 & 3.1\,${\scriptstyle \mathsf{E}}$-11$\!\!\!$\tabularnewline
$\!\!\!$Accur Aver  & 1.9\,${\scriptstyle \mathsf{E}}$-12 & 1.2\,${\scriptstyle \mathsf{E}}$-12 & 1.4\,${\scriptstyle \mathsf{E}}$-12 & 9.4\,${\scriptstyle \mathsf{E}}$-12 & 1.1\,${\scriptstyle \mathsf{E}}$-11 & 6.9\,${\scriptstyle \mathsf{E}}$-12 & 1.8\,${\scriptstyle \mathsf{E}}$-11$\!\!\!$\tabularnewline
$\!\!\!$Accur Aver Stab & 3.8\,${\scriptstyle \mathsf{E}}$-15 & 1.8\,${\scriptstyle \mathsf{E}}$-15 & 2.0\,${\scriptstyle \mathsf{E}}$-15 & 1.9\,${\scriptstyle \mathsf{E}}$-15 & 1.9\,${\scriptstyle \mathsf{E}}$-15 & 2.5\,${\scriptstyle \mathsf{E}}$-15 & 1.9\,${\scriptstyle \mathsf{E}}$-15$\!\!\!$\tabularnewline
\bottomrule
\end{tabular}
\par\end{centering}
\centering{}\caption{1D, bandwidth 15\%\label{tab:1DKDE0.15}}
\end{table}

\begin{table}[H]
\begin{centering}
\begin{tabular}{llllllll}
\toprule 
$\!\!\!$Nb particles & 20,000 & 40,000 & 80,000 & 160,000 & 320,000 & 640,000 & 1,280,000$\!\!\!$\tabularnewline
\midrule
$\!\!\!$Fast kernel time & 0.00 & 0.01 & 0.02 & 0.05 & 0.09 & 0.20 & 0.41\tabularnewline
$\!\!\!$Naive time  & 4.20 & 17 & 67 & 270 & 1,100 & 4,300 & 17,000\tabularnewline
\midrule 
$\!\!\!$Accur Worst & 2.1\,${\scriptstyle \mathsf{E}}$-10 & 7.2\,${\scriptstyle \mathsf{E}}$-11 & 2.8\,${\scriptstyle \mathsf{E}}$-10 & 7.8\,${\scriptstyle \mathsf{E}}$-09 & 3.6\,${\scriptstyle \mathsf{E}}$-08 & 2.9\,${\scriptstyle \mathsf{E}}$-07 & 1.3\,${\scriptstyle \mathsf{E}}$-07$\!\!\!$\tabularnewline
$\!\!\!$Accur Worst Stab  & 4.2\,${\scriptstyle \mathsf{E}}$-13 & 6.4\,${\scriptstyle \mathsf{E}}$-13 & 1.7\,${\scriptstyle \mathsf{E}}$-12 & 3.5\,${\scriptstyle \mathsf{E}}$-12 & 1.3\,${\scriptstyle \mathsf{E}}$-11 & 3.9\,${\scriptstyle \mathsf{E}}$-11 & 3.1\,${\scriptstyle \mathsf{E}}$-11$\!\!\!$\tabularnewline
$\!\!\!$Accur Aver  & 2.3\,${\scriptstyle \mathsf{E}}$-13 & 8.1\,${\scriptstyle \mathsf{E}}$-14 & 9.2\,${\scriptstyle \mathsf{E}}$-14 & 1.2\,${\scriptstyle \mathsf{E}}$-12 & 2.3\,${\scriptstyle \mathsf{E}}$-12 & 2.0\,${\scriptstyle \mathsf{E}}$-12 & 3.3\,${\scriptstyle \mathsf{E}}$-12$\!\!\!$\tabularnewline
$\!\!\!$Accur Aver Stab & 7.6\,${\scriptstyle \mathsf{E}}$-16 & 8.0\,${\scriptstyle \mathsf{E}}$-16 & 8.1\,${\scriptstyle \mathsf{E}}$-16 & 8.7\,${\scriptstyle \mathsf{E}}$-16 & 8.8\,${\scriptstyle \mathsf{E}}$-16 & 9.0\,${\scriptstyle \mathsf{E}}$-16 & 8.9\,${\scriptstyle \mathsf{E}}$-16$\!\!\!$\tabularnewline
\bottomrule
\end{tabular}
\par\end{centering}
\centering{}\caption{1D, bandwidth 25\%\label{tab:1DKDE0.25}}
\end{table}

The runtime of the fast summation algorithm is nearly independent
of the size of the bandwidth. It is not the case for the naive implementation.
Indeed the larger the bandwidth, the more input points contribute
to the kernel summation \eqref{eq:localdens} resulting in more operations
for larger bandwidths. This independence with respect to bandwidth
size is another advantage of the fast sum updating approach over alternative
methods such as naive summation or dual-tree methods.

\begin{figure}[H]
\begin{minipage}[t]{0.48\columnwidth}%
\includegraphics[width=0.38\paperwidth]{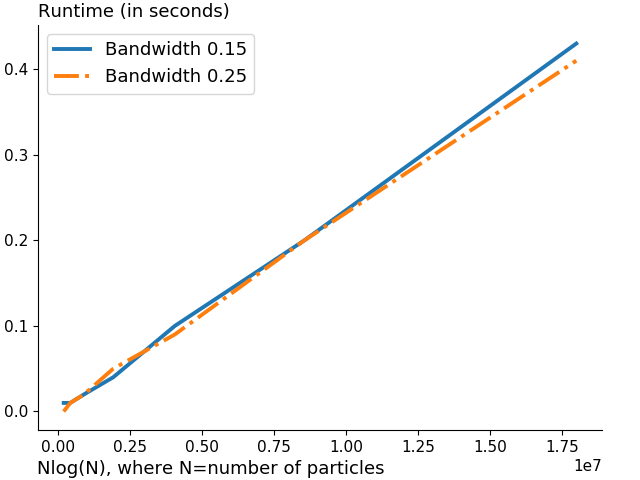}%
\end{minipage}\hfill{}%
\begin{minipage}[t]{0.48\columnwidth}%
\includegraphics[width=0.38\paperwidth]{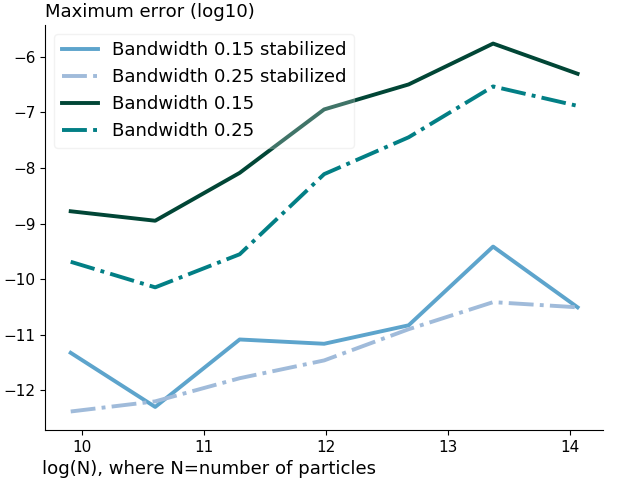}%
\end{minipage}

\caption{Speed and accuracy of fast univariate kernel summation\label{fig:Fast1DKDE}}
\end{figure}

Figure \ref{fig:Fast1DKDE} (left-hand side) clearly demonstrates
that the computational time is in $N\log N$ as expected (subsection
\ref{subsec:Complexity}), while the right-hand side clearly demonstrates
the efficiency of the stabilization. As expected, the cumulative float-rounding
error slowly grows with the sample size $N$, but remains negligible
even on the largest sample sizes. 

\subsubsection{Multivariate case}

Tables \ref{tab:2DKDE0.15} and \ref{tab:2DKDE0.25} report our speed
and accuracy results in the bivariate case. Once again, the fast summation
algorithm is vastly faster than naive summation (less than one second
versus more than seven hours for $1,28$ million points for example),
and the runtime of the fast summation algorithm is independent of
the size of the bandwidth. Moreover, we observe a very good accuracy
(much better than the univariate case for example), even without using
any summation stabilization algorithm (subsection \ref{subsec:Numerical-stability}).

\begin{table}[H]
\begin{centering}
\begin{tabular}{llllllll}
\toprule 
$\!\!\!$Nb particles & 20,000 & 40,000 & 80,000 & 160,000 & 320,000 & 640,000 & 1,280,000$\!\!\!$\tabularnewline
\midrule
$\!\!\!$Fast kernel time & 0.02 & 0.02 & 0.04 & 0.09 & 0.20 & 0.43 & 0.89\tabularnewline
$\!\!\!$Naive time  & 6.50 & 26 & 100 & 420 & 1,700 & 6,700 & 27,000\tabularnewline
\midrule 
$\!\!\!$Accur Worst & 3.2\,${\scriptstyle \mathsf{E}}$-12 & 1.9\,${\scriptstyle \mathsf{E}}$-12 & 3.0\,${\scriptstyle \mathsf{E}}$-10 & 4.5\,${\scriptstyle \mathsf{E}}$-10 & 4.0\,${\scriptstyle \mathsf{E}}$-11 & 7.2\,${\scriptstyle \mathsf{E}}$-08 & 3.5\,${\scriptstyle \mathsf{E}}$-09$\!\!\!$\tabularnewline
$\!\!\!$Accur Worst Stab  & 4.4\,${\scriptstyle \mathsf{E}}$-13 & 1.6\,${\scriptstyle \mathsf{E}}$-13 & 3.3\,${\scriptstyle \mathsf{E}}$-12 & 1.7\,${\scriptstyle \mathsf{E}}$-11 & 4.1\,${\scriptstyle \mathsf{E}}$-13 & 1.1\,${\scriptstyle \mathsf{E}}$-10 & 3.0\,${\scriptstyle \mathsf{E}}$-11$\!\!\!$\tabularnewline
$\!\!\!$Accur Aver  & 8.3\,${\scriptstyle \mathsf{E}}$-15 & 3.2\,${\scriptstyle \mathsf{E}}$-15 & 2.3\,${\scriptstyle \mathsf{E}}$-14 & 8.0\,${\scriptstyle \mathsf{E}}$-15 & 1.8\,${\scriptstyle \mathsf{E}}$-14 & 1.8\,${\scriptstyle \mathsf{E}}$-13 & 5.3\,${\scriptstyle \mathsf{E}}$-14$\!\!\!$\tabularnewline
$\!\!\!$Accur Aver Stab & 3.7\,${\scriptstyle \mathsf{E}}$-16 & 3.0\,${\scriptstyle \mathsf{E}}$-16 & 4.5\,${\scriptstyle \mathsf{E}}$-16 & 4.9\,${\scriptstyle \mathsf{E}}$-16 & 3.0\,${\scriptstyle \mathsf{E}}$-16 & 6.4\,${\scriptstyle \mathsf{E}}$-16 & 4.3\,${\scriptstyle \mathsf{E}}$-16$\!\!\!$\tabularnewline
\bottomrule
\end{tabular}
\par\end{centering}
\centering{}\caption{2D, bandwidth 15\%\label{tab:2DKDE0.15}}
\end{table}
\begin{table}[H]
\begin{centering}
\begin{tabular}{llllllll}
\toprule 
$\!\!\!$Nb particles & 20,000 & 40,000 & 80,000 & 160,000 & 320,000 & 640,000 & 1,280,000$\!\!\!$\tabularnewline
\midrule
$\!\!\!$Fast kernel time & 0.01 & 0.02 & 0.04 & 0.08 & 0.19 & 0.41 & 0.88\tabularnewline
$\!\!\!$Naive time  & 8.30 & 33 & 130 & 540 & 2,100 & 8,700 & 34,000\tabularnewline
\midrule 
$\!\!\!$Accur Worst & 1.8\,${\scriptstyle \mathsf{E}}$-11 & 6.0\,${\scriptstyle \mathsf{E}}$-12 & 6.9\,${\scriptstyle \mathsf{E}}$-11 & 8.3\,${\scriptstyle \mathsf{E}}$-11 & 1.6\,${\scriptstyle \mathsf{E}}$-09 & 1.1\,${\scriptstyle \mathsf{E}}$-09 & 4.2\,${\scriptstyle \mathsf{E}}$-09$\!\!\!$\tabularnewline
$\!\!\!$Accur Worst Stab  & 6.5\,${\scriptstyle \mathsf{E}}$-13 & 4.2\,${\scriptstyle \mathsf{E}}$-13 & 2.7\,${\scriptstyle \mathsf{E}}$-13 & 1.9\,${\scriptstyle \mathsf{E}}$-11 & 8.2\,${\scriptstyle \mathsf{E}}$-12 & 9.7\,${\scriptstyle \mathsf{E}}$-12 & 3.4\,${\scriptstyle \mathsf{E}}$-11$\!\!\!$\tabularnewline
$\!\!\!$Accur Aver  & 8.5\,${\scriptstyle \mathsf{E}}$-15 & 3.8\,${\scriptstyle \mathsf{E}}$-15 & 8.7\,${\scriptstyle \mathsf{E}}$-15 & 4.6\,${\scriptstyle \mathsf{E}}$-15 & 2.7\,${\scriptstyle \mathsf{E}}$-14 & 2.1\,${\scriptstyle \mathsf{E}}$-14 & 4.8\,${\scriptstyle \mathsf{E}}$-14$\!\!\!$\tabularnewline
$\!\!\!$Accur Aver Stab & 3.3\,${\scriptstyle \mathsf{E}}$-16 & 3.1\,${\scriptstyle \mathsf{E}}$-16 & 2.7\,${\scriptstyle \mathsf{E}}$-16 & 6.7\,${\scriptstyle \mathsf{E}}$-16 & 4.4\,${\scriptstyle \mathsf{E}}$-16 & 2.9\,${\scriptstyle \mathsf{E}}$-16 & 5.1\,${\scriptstyle \mathsf{E}}$-16$\!\!\!$\tabularnewline
\bottomrule
\end{tabular}
\par\end{centering}
\centering{}\caption{2D, bandwidth 25\%\label{tab:2DKDE0.25}}
\end{table}
\begin{figure}[H]
\begin{minipage}[t]{0.48\columnwidth}%
\includegraphics[width=0.38\paperwidth]{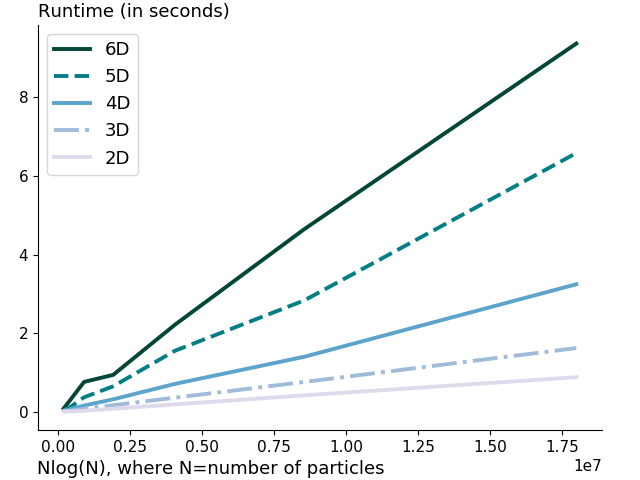}%
\end{minipage}\hfill{}%
\begin{minipage}[t]{0.48\columnwidth}%
\includegraphics[width=0.38\paperwidth]{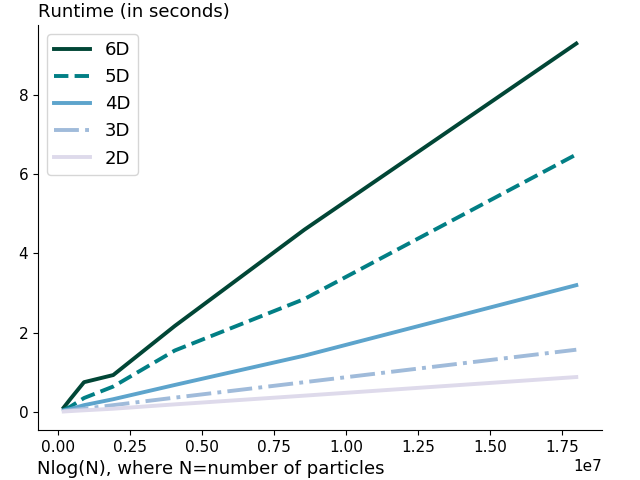}%
\end{minipage}

\caption{Runtime of fast kernel summation (left: bandwidth 15\%; right: bandwidth
25\%)\label{fig:timeFastKDE}}
\end{figure}

\begin{figure}[H]
\begin{minipage}[t]{0.48\columnwidth}%
\includegraphics[width=0.38\paperwidth]{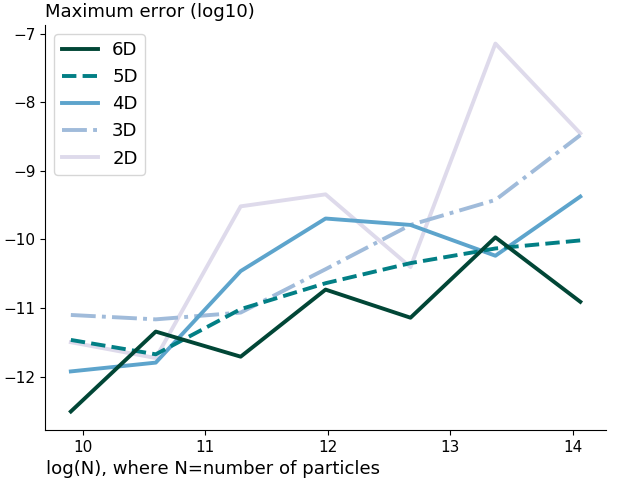}%
\end{minipage}\hfill{}%
\begin{minipage}[t]{0.48\columnwidth}%
\includegraphics[width=0.38\paperwidth]{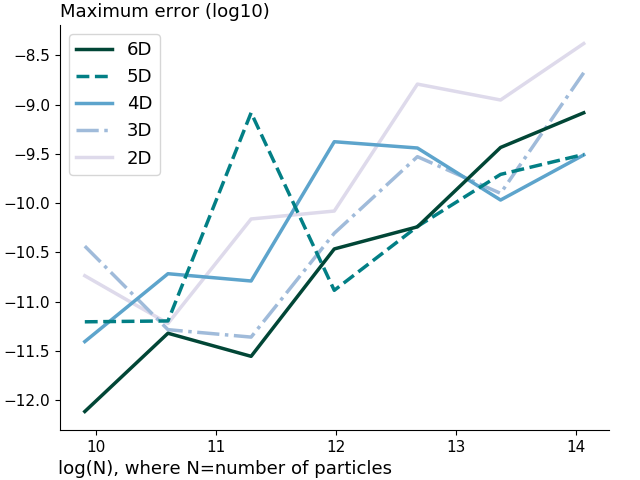}%
\end{minipage}

\caption{$\log$10 of maximum relative error w.r.t. $\log$N (left: bandwidth
15\%; right: 25\%)\label{fig:accurDimKDE}}
\end{figure}

Figures \ref{fig:timeFastKDE} and \ref{fig:accurDimKDE} report multidimensional
results up to dimension $6$. Figure \ref{fig:timeFastKDE} demonstrates
once again that the computational runtime is clearly in $N\log N$,
while Figure \ref{fig:accurDimKDE} shows that the accuracy is very
good, even without summation stabilization.

\subsection{Fast locally linear regression\label{subsec:klin-test}}

For comprehensiveness, we now verify that our numerical observations
from subsection \ref{subsec:kde-test} still hold for the harder locally
linear regression problem (equations \eqref{eq:localreg1} and \eqref{eq:localreg1-d}).

\subsubsection{Univariate case}

Once again, we first consider the univariate case. Tables \ref{tab:1D015}
and \ref{tab:1D025} summarize the results obtained by Algorithm \ref{alg:fast1Dkreg}
with the two different bandwidths. The results are very similar to
the kernel density estimation case.

\begin{figure}[H]
\begin{minipage}[t]{0.48\columnwidth}%
\includegraphics[width=0.38\paperwidth]{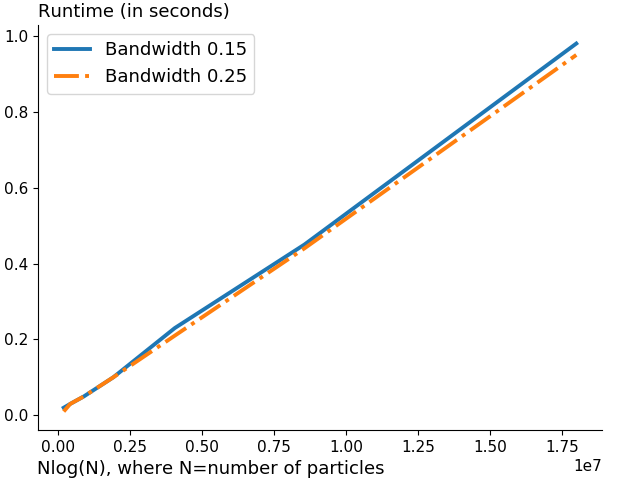}%
\end{minipage}\hfill{}%
\begin{minipage}[t]{0.48\columnwidth}%
\includegraphics[width=0.38\paperwidth]{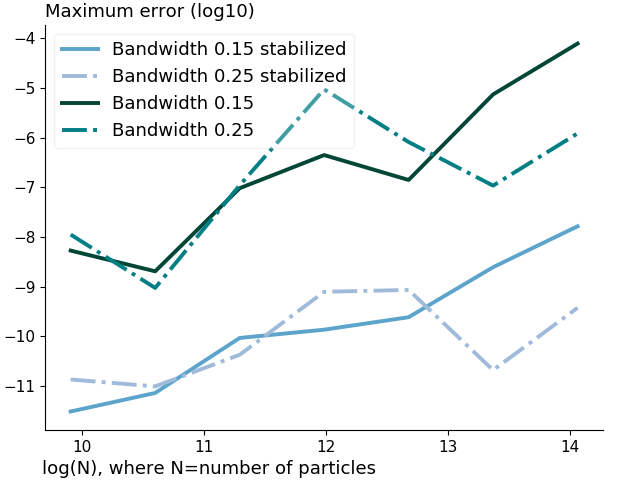}%
\end{minipage}

\caption{Speed and accuracy of fast univariate kernel summation\label{fig:Fast1D}}
\end{figure}

Figure \ref{fig:Fast1D} demonstrates that, as in the kernel density
estimation case, the computational runtime is clearly in $N\log N$
and that the simple summation stabilization we implemented is very
effective (the numerical accuracy is improved by a factor $1,500$
on average).

\begin{table}[H]
\begin{centering}
\begin{tabular}{llllllll}
\toprule 
$\!\!\!$Nb particles & 20,000 & 40,000 & 80,000 & 160,000 & 320,000 & 640,000 & 1,280,000$\!\!\!$\tabularnewline
\midrule
$\!\!\!$Fast kernel time & 0.02 & 0.03 & 0.05 & 0.1 & 0.23 & 0.45 & 0.98\tabularnewline
$\!\!\!$Naive time  & 4.50 & 18 & 71 & 280  & 1,100 & 4,500 & 18,000\tabularnewline
\midrule 
$\!\!\!$Accur Worst & 5.3\,${\scriptstyle \mathsf{E}}$-09 & 2.0\,${\scriptstyle \mathsf{E}}$-09 & 9.6\,${\scriptstyle \mathsf{E}}$-08 & 4.4\,${\scriptstyle \mathsf{E}}$-07 & 1.4\,${\scriptstyle \mathsf{E}}$-07 & 7.4\,${\scriptstyle \mathsf{E}}$-06 & 7.8\,${\scriptstyle \mathsf{E}}$-05$\!\!\!$\tabularnewline
$\!\!\!$Accur Worst Stab  & 3.1\,${\scriptstyle \mathsf{E}}$-12 & 7.2\,${\scriptstyle \mathsf{E}}$-12 & 9.2\,${\scriptstyle \mathsf{E}}$-11 & 1.4\,${\scriptstyle \mathsf{E}}$-10 & 2.4\,${\scriptstyle \mathsf{E}}$-10 & 2.5\,${\scriptstyle \mathsf{E}}$-09 & 1.6\,${\scriptstyle \mathsf{E}}$-08$\!\!\!$\tabularnewline
$\!\!\!$Accur Aver  & 4.4\,${\scriptstyle \mathsf{E}}$-12 & 2.1\,${\scriptstyle \mathsf{E}}$-12 & 9.1\,${\scriptstyle \mathsf{E}}$-12 & 1.4\,${\scriptstyle \mathsf{E}}$-11 & 6.1\,${\scriptstyle \mathsf{E}}$-12 & 2.9\,${\scriptstyle \mathsf{E}}$-11 & 8.5\,${\scriptstyle \mathsf{E}}$-11$\!\!\!$\tabularnewline
$\!\!\!$Accur Aver Stab & 5.2\,${\scriptstyle \mathsf{E}}$-15 & 5.5\,${\scriptstyle \mathsf{E}}$-15 & 6.2\,${\scriptstyle \mathsf{E}}$-15 & 5.8\,${\scriptstyle \mathsf{E}}$-15 & 6.4\,${\scriptstyle \mathsf{E}}$-15 & 9.5\,${\scriptstyle \mathsf{E}}$-15 & 2.1\,${\scriptstyle \mathsf{E}}$-14$\!\!\!$\tabularnewline
\bottomrule
\end{tabular}
\par\end{centering}
\centering{}\caption{1D results, bandwidth 15\%\label{tab:1D015}}
\end{table}

\begin{table}[H]
\begin{centering}
\begin{tabular}{llllllll}
\toprule 
$\!\!\!$Nb particles & 20,000 & 40,000 & 80,000 & 160,000 & 320,000 & 640,000 & 1,280,000$\!\!\!$\tabularnewline
\midrule
$\!\!\!$Fast kernel time & 0.01  & 0.03  & 0.05  & 0.1  & 0.21  & 0.44  & 0.95 \tabularnewline
$\!\!\!$Naive time  & 6.60  & 26  & 100 & 420  & 1,700  & 6,700  & 27,000 \tabularnewline
\midrule
$\!\!\!$Accur Worst & 1.1\,${\scriptstyle \mathsf{E}}$-08 & 9.4\,${\scriptstyle \mathsf{E}}$-10 & 1.1\,${\scriptstyle \mathsf{E}}$-07 & 9.4\,${\scriptstyle \mathsf{E}}$-06 & 8.1\,${\scriptstyle \mathsf{E}}$-07 & 1.1\,${\scriptstyle \mathsf{E}}$-07 & 1.2\,${\scriptstyle \mathsf{E}}$-06$\!\!\!$\tabularnewline
$\!\!\!$Accur Worst Stab  & 1.3\,${\scriptstyle \mathsf{E}}$-11 & 9.8\,${\scriptstyle \mathsf{E}}$-12 & 4.2\,${\scriptstyle \mathsf{E}}$-11 & 7.8\,${\scriptstyle \mathsf{E}}$-10 & 8.5\,${\scriptstyle \mathsf{E}}$-10 & 2.1\,${\scriptstyle \mathsf{E}}$-11 & 3.7\,${\scriptstyle \mathsf{E}}$-10$\!\!\!$\tabularnewline
$\!\!\!$Accur Aver  & 2.3\,${\scriptstyle \mathsf{E}}$-12 & 6.6\,${\scriptstyle \mathsf{E}}$-13 & 4.3\,${\scriptstyle \mathsf{E}}$-12 & 6.9\,${\scriptstyle \mathsf{E}}$-11 & 5.8\,${\scriptstyle \mathsf{E}}$-12 & 4.6\,${\scriptstyle \mathsf{E}}$-12 & 9.4\,${\scriptstyle \mathsf{E}}$-12$\!\!\!$\tabularnewline
$\!\!\!$Accur Aver Stab & 2.7\,${\scriptstyle \mathsf{E}}$-15 & 2.2\,${\scriptstyle \mathsf{E}}$-15 & 2.3\,${\scriptstyle \mathsf{E}}$-15 & 8.6\,${\scriptstyle \mathsf{E}}$-15 & 4.7\,${\scriptstyle \mathsf{E}}$-15 & 1.9\,${\scriptstyle \mathsf{E}}$-15 & 2.3\,${\scriptstyle \mathsf{E}}$-15$\!\!\!$\tabularnewline
\bottomrule
\end{tabular}
\par\end{centering}
\centering{}\caption{1D results, bandwidth 25\%\label{tab:1D025}}
\end{table}

\subsubsection{Multivariate case}

Tables \ref{tab:2D015} and \ref{tab:2D025} report our speed and
accuracy results in the bivariate locally linear regression case.
The results are once again qualitatively very similar to the kernel
density estimation case.

\begin{table}[H]
\begin{centering}
\begin{tabular}{llllllll}
\toprule 
$\!\!\!$Nb particles & 20,000 & 40,000 & 80,000 & 160,000 & 320,000 & 640,000 & 1,280,000$\!\!\!$\tabularnewline
\midrule
$\!\!\!$Fast kernel time & 0.03  & 0.06  & 0.12  & 0.26  & 0.52  & 1.1  & 2.22 \tabularnewline
$\!\!\!$Naive time  & 9.80  & 40  & 160 & 630 & 2,500 & 10,000 & 40,000\tabularnewline
\midrule
$\!\!\!$Accur Worst & 8.6\,${\scriptstyle \mathsf{E}}$-11  & 2.3\,${\scriptstyle \mathsf{E}}$-11 & 5.6\,${\scriptstyle \mathsf{E}}$-10 & 4.9\,${\scriptstyle \mathsf{E}}$-10 & 2.6\,${\scriptstyle \mathsf{E}}$-09 & 2.7\,${\scriptstyle \mathsf{E}}$-09 & 4.9\,${\scriptstyle \mathsf{E}}$-09$\!\!\!$\tabularnewline
$\!\!\!$Accur Aver  & 5.3\,${\scriptstyle \mathsf{E}}$-14 & 1.1\,${\scriptstyle \mathsf{E}}$-14 & 5.1\,${\scriptstyle \mathsf{E}}$-14 & 2.4\,${\scriptstyle \mathsf{E}}$-14 & 7.1\,${\scriptstyle \mathsf{E}}$-14 & 9.2\,${\scriptstyle \mathsf{E}}$-14 & 1.3\,${\scriptstyle \mathsf{E}}$-13$\!\!\!$\tabularnewline
\bottomrule
\end{tabular}
\par\end{centering}
\centering{}\caption{2D results, bandwidth 15\%\label{tab:2D015}}
\end{table}

\begin{table}[H]
\begin{centering}
\begin{tabular}{llllllll}
\hline 
$\!\!\!$Nb particles & 20,000 & 40,000 & 80,000 & 160,000 & 320,000 & 640,000 & 1,280,000$\!\!\!$\tabularnewline
\hline 
$\!\!\!$Fast kernel time & 0.02  & 0.06  & 0.13  & 0.25  & 0.51  & 1.06  & 2.17 \tabularnewline
$\!\!\!$Naive time  & 14  & 54  & 220 & 870 & 3,500 & 14,000 & 56,000\tabularnewline
\hline 
$\!\!\!$Accur Worst & 9.5\,${\scriptstyle \mathsf{E}}$-11 & 7.6\,${\scriptstyle \mathsf{E}}$-11 & 8.0\,${\scriptstyle \mathsf{E}}$-11 & 5.7\,${\scriptstyle \mathsf{E}}$-11 & 2.3\,${\scriptstyle \mathsf{E}}$-09 & 2.5\,${\scriptstyle \mathsf{E}}$-09 & 3.5\,${\scriptstyle \mathsf{E}}$-09$\!\!\!$\tabularnewline
$\!\!\!$Accur Aver  & 1.7\,${\scriptstyle \mathsf{E}}$-14 & 6.3\,${\scriptstyle \mathsf{E}}$-15 & 2.4\,${\scriptstyle \mathsf{E}}$-14 & 7.3\,${\scriptstyle \mathsf{E}}$-15 & 5.9\,${\scriptstyle \mathsf{E}}$-14 & 7.2\,${\scriptstyle \mathsf{E}}$-14 & 5.0\,${\scriptstyle \mathsf{E}}$-14$\!\!\!$\tabularnewline
\hline 
\end{tabular}
\par\end{centering}
\centering{}\caption{2D results, bandwidth 25\%\label{tab:2D025}}
\end{table}

\begin{figure}[H]
\begin{minipage}[t]{0.48\columnwidth}%
\includegraphics[width=0.38\paperwidth]{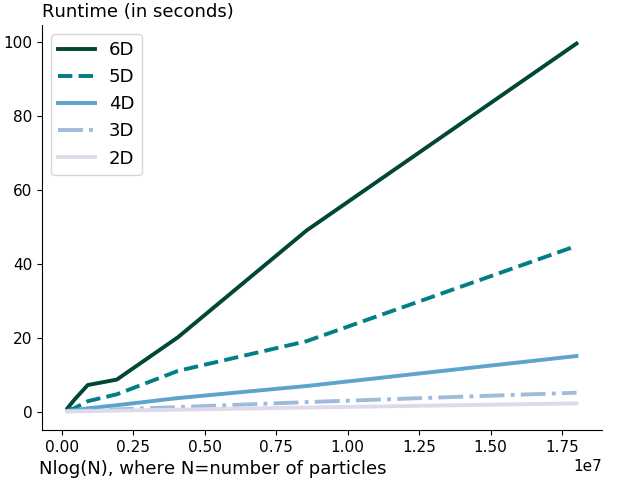}%
\end{minipage}\hfill{}%
\begin{minipage}[t]{0.48\columnwidth}%
\includegraphics[width=0.38\paperwidth]{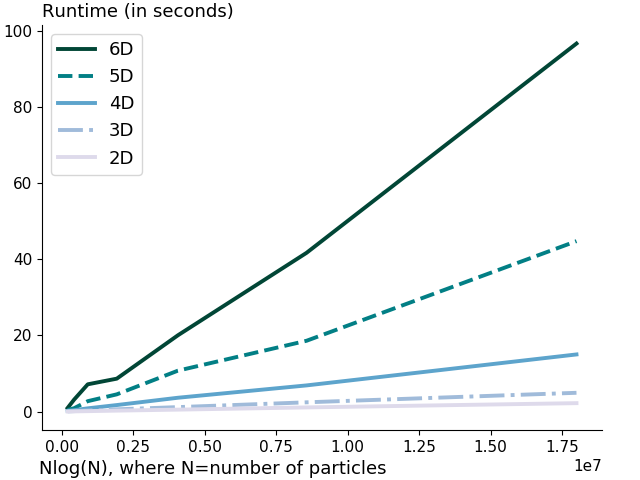}%
\end{minipage}

\caption{Runtime of fast kernel summation (left: bandwidth 15\%; right: bandwidth
25\%)\label{fig:timeFast} }
\end{figure}

\begin{figure}[H]
\begin{minipage}[t]{0.48\columnwidth}%
\includegraphics[width=0.38\paperwidth]{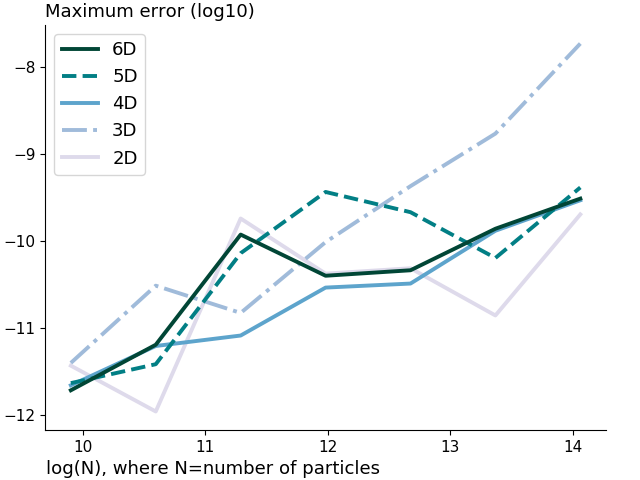}%
\end{minipage}\hfill{}%
\begin{minipage}[t]{0.48\columnwidth}%
\includegraphics[width=0.38\paperwidth]{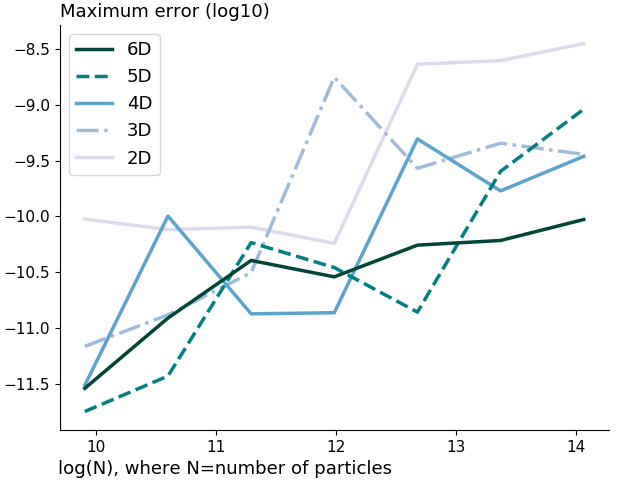}%
\end{minipage}

\caption{$\log10$ of maximum relative error w.r.t. $\log\mathrm{N}$ (left:
bandwidth 15\%; right: 25\%)\label{fig:accurDim}}
\end{figure}

Finally, Figures \ref{fig:timeFast} and \ref{fig:accurDim} report
multivariate locally linear regression results up to dimension $6$,
demonstrating once again the $N\log N$ computational complexity and
the very good accuracy. Note however that compared to the kernel density
estimation case, the runtime grows much more quickly with the dimension
of the problem. This is due to the higher number of terms to track
to perform the locally linear regressions \eqref{eq:localreg1-d-explicit}
compared to one single kernel density estimation. 

\section{Conclusion}

Fast and exact kernel density estimation can be achieved by the \textit{fast
sum updating} algorithm (\citet{Gasser89}, \citet{Seifert94}). With
$N$ input points drawn from the density to estimate, and $M$ evaluation
points where this density needs to be estimated, the fast sum updating
algorithm requires $\mathcal{O}(M\log M+N\log N)$ operations, which
is a vast improvement over the $\mathcal{O}(MN)$ operations required
by direct kernel summation. This paper revisits the fast sum updating
algorithm and extends it in several ways. 

The main contribution is the extension, for the first time, of the
fast sum updating algorithm to the general multivariate case, opening
the door to a vast class of practical density estimation and regression
problems. The original concern in \citet{Seifert94} with floating-point
summation instability due to float-rounding errors can be completely
addressed by the use of exact floating-point summation algorithms.
Our numerical tests show that the cumulative float-rounding error
is already negligible when using double-precision floats (in line
with \citet{Fan94}), and that very simple compensated summation algorithms
such as the M\o ller-Kahan algorithm can already bring significant
accuracy improvements. 

In addition, we show that fast sum updating is compatible with a larger
list of kernels, including the triangular kernel, the Silverman kernel,
the cosine kernel, and the newly introduced hyperbolic cosine kernel,
than what was usually assumed in the literature. We introduce the
multivariate additive kernel, which greatly improves the speed of
fast sum updating in high dimension compared to product kernels. Importantly,
we describe how fast sum updating is compatible with balloon adaptive
bandwidths, and propose a fast approximate k-nearest-neighbor algorithm
for the adaptive bandwidth. 

The proposed multivariate extension does not impose any restriction
on the input or output samples, but does require the evaluation points
to lie on a possibly non-uniform grid. We describe how to prerotate
the input data and construct a suitable grid to ease the interpolation
of density estimates to any evaluation sample by multilinear interpolation
or fast inverse distance weighting.

Our multivariate kernel density and locally linear regression tests
confirm numerically the vastly improved computational speed compared
to naive kernel summation, as well as the accuracy and stability of
the method. A natural area for future research would be to examine
density estimation or regression applications for which computational
speed is a major issue. It would in particular be worth investigating
how this algorithm compares, in terms of speed and accuracy, to alternative
fast but approximate density estimation algorithms such as the Fast
Fourier Transform with binning or the Fast Gauss Transform.

\subsubsection*{Acknowledgements}

The authors are grateful to the anonymous referees for their valuable
comments. Xavier Warin acknowledges the financial support of ANR project
CAESARS (ANR-15-CE05-0024).

\clearpage{}
\bibliographystyle{chicago}
\bibliography{Biblio}
\clearpage{}

\appendix

\section{Kernels compatible with fast sum updating\label{sec:fast-sum-kernels}}

This Appendix details how to implement the fast sum updating algorithm
for the kernels listed in Table \ref{tab:fast-sum-kernels}. Three
classes of kernels admit the type of separation between sources and
targets required for the fast sum updating algorithm: polynomial kernels
(subsection \ref{subsec:Polynomial-kernels}), absolute kernels (subsection
\ref{subsec:Absolute-kernels}) and cosine kernels (subsection \ref{subsec:Cosine-kernels}).
In addition, fast sum updating is still applicable to kernels which
combine features from these three classes (subsection \ref{subsec:Combinations}).
In the literature, \citet{Seifert94} covered the case of polynomial
kernels, while \citet{Chen06} covered the Laplacian kernel. The present
paper extends the applicability of fast sum updating to the triangular
kernel, cosine kernel, hyperbolic cosine kernel, and combinations
such as the tricube and Silverman kernels.

Specifically, we detail how to decompose the sums
\[
\frac{1}{N}\sum_{i=1}^{N}K_{h}(x_{i}-z_{j})x_{i}^{p}y_{i}^{q}=\frac{1}{Nh_{j}}\sum_{i=1}^{N}K\left(\frac{x_{i}-z_{j}}{h}\right)x_{i}^{p}y_{i}^{q}\,,\,\,j\in\{1,2,\ldots,M\}
\]
into fast updatable sums of the type
\begin{align}
\mathcal{S}^{p,q}\left(f,[L,R]\right): & =\sum_{i=1}^{N}f(x_{i})x_{i}^{p}y_{i}^{q}\mathbbm{1}\{L\leq x_{i}\leq R\}\label{eq:SfpqLR}
\end{align}
Equation \eqref{eq:SfpqLR} is a generalization of the sum \eqref{eq:SpqLR}
used in Section \ref{sec:Fast-sum-updating} for the Epanechnikov
kernel. The additional $f(x_{i})$ term in the sum is necessary for
such kernels as the cosine or Laplacian ones.

Whenever possible, we will use adaptive kernels $h=h_{j}$ (balloon
estimator, cf. subsection \ref{subsec:knn-bandwidth}). Some kernels,
such as polynomial kernels, can combine adaptive bandwidths with fast
sum updating, but some other kernels cannot, as explained in the subsections
below.

\subsection{Polynomial kernels\label{subsec:Polynomial-kernels}}

The class of polynomial kernels, in particular the class of symmetric
beta kernels 
\[
K(u)=\frac{(1-u^{2})^{\alpha}}{2^{2\alpha+1}\frac{\Gamma(\alpha+1)\Gamma(\alpha+1)}{\Gamma(2\alpha+2)}}\mathbbm{1}\{\left|u\right|\leq1\}
\]
includes several classical kernels: the uniform/rectangular kernel
($\alpha=0$), the Epanechnikov/parabolic kernel ($\alpha=1$), the
quartic/biweight kernel ($\alpha=2$) and the triweight kernel ($\alpha=3$).
We recall from Section \ref{sec:Fast-sum-updating} how to decompose
the Epanechnikov kernel $K(u)=\frac{3}{4}(1-u^{2})\mathbbm{1}\{\left|u\right|\leq1\}$.
By expanding the square term:
\begin{align*}
 & \sum_{i=1}^{N}K\!\left(\frac{x_{i}-z_{j}}{h_{j}}\right)x_{i}^{p}y_{i}^{q}\mathbbm{1}\left\{ \left|\frac{x_{i}-z_{j}}{h_{j}}\right|\leq1\right\} \\
 & =\frac{3}{4}\sum_{i=1}^{N}\left\{ \left(1-\frac{z_{j}^{2}}{h_{j}^{2}}\right)+\frac{2z_{j}}{h_{j}^{2}}x_{i}-\frac{1}{h_{j}^{2}}x_{i}^{2}\right\} x_{i}^{p}y_{i}^{q}\mathbbm{1}\left\{ z_{j}\!-\!h_{j}\leq x_{i}\leq z_{j}\!+\!h_{j}\right\} \\
 & =\frac{3}{4}\left(1-\frac{z_{j}^{2}}{h_{j}^{2}}\right)\mathcal{S}^{p,q}\left(1,[z_{j}\!-\!h_{j},z_{j}\!+\!h_{j}]\right)+\frac{3}{4}\frac{2z_{j}}{h_{j}^{2}}\mathcal{S}^{p+1,q}\left(1,[z_{j}\!-\!h_{j},z_{j}\!+\!h_{j}]\right)\\
 & -\frac{3}{4}\frac{1}{h_{j}^{2}}\mathcal{S}^{p+2,q}\left(1,[z_{j}\!-\!h_{j},z_{j}\!+\!h_{j}]\right)
\end{align*}
 The other kernels within this class can be decomposed in a similar
manner by expanding the power terms. 

\subsection{Absolute kernels\label{subsec:Absolute-kernels}}

The class of absolute kernels contains kernels based on the absolute
value $\left|u\right|$, such as the triangular kernel and the Laplacian
kernel.

For the triangular kernel, $K(u)=(1-\left|u\right|)\mathbbm{1}\{\left|u\right|\leq1\}$
and
\begin{align*}
 & \sum_{i=1}^{N}K\!\left(\frac{x_{i}-z_{j}}{h_{j}}\right)x_{i}^{p}y_{i}^{q}\mathbbm{1}\left\{ \left|\frac{x_{i}-z_{j}}{h_{j}}\right|\leq1\right\} \\
 & =\sum_{i=1}^{N}\left(1-\frac{x_{i}-z_{j}}{h_{j}}\right)x_{i}^{p}y_{i}^{q}\mathbbm{1}\left\{ z_{j}\leq x_{i}\leq z_{j}\!+\!h_{j}\right\} +\sum_{i=1}^{N}\left(1-\frac{z_{j}-x_{i}}{h_{j}}\right)x_{i}^{p}y_{i}^{q}\mathbbm{1}\left\{ z_{j}\!-\!h_{j}\leq x_{i}<z_{j}\right\} \\
 & =\left(1+\frac{z_{j}}{h_{j}}\right)\mathcal{S}^{p,q}\left(1,[z_{j},z_{j}\!+\!h_{j}]\right)-\frac{1}{h_{j}}\mathcal{S}^{p+1,q}\left(1,[z_{j},z_{j}\!+\!h_{j}]\right)\\
 & +\left(1-\frac{z_{j}}{h_{j}}\right)\mathcal{S}^{p,q}\left(1,[z_{j}\!-\!h_{j},z_{j}[\right)+\frac{1}{h_{j}}\mathcal{S}^{p+1,q}\left(1,[z_{j}\!-\!h_{j},z_{j}[\right)
\end{align*}

For the Laplacian kernel, $K(u)=\frac{1}{2}\exp(-\left|u\right|)$
and
\begin{align*}
 & \sum_{i=1}^{N}K\!\left(\frac{x_{i}-z_{j}}{h}\right)x_{i}^{p}y_{i}^{q}\\
 & =\frac{1}{2}\sum_{i=1}^{N}\exp\!\left(-\frac{x_{i}\!-\!z_{j}}{h}\right)x_{i}^{p}y_{i}^{q}\mathbbm{1}\left\{ z_{j}\leq x_{i}\right\} +\frac{1}{2}\sum_{i=1}^{N}\exp\!\left(-\frac{z_{j}\!-\!x_{i}}{h}\right)x_{i}^{p}y_{i}^{q}\mathbbm{1}\left\{ x_{i}<z_{j}\right\} \\
 & =\frac{1}{2}\exp\left(\frac{z_{j}}{h}\right)\mathcal{S}^{p,q}\left(\exp(-./h),[z_{j},\infty[\right)+\frac{1}{2}\exp\left(-\frac{z_{j}}{h}\right)\mathcal{S}^{p,q}\left(\exp(./h),]-\infty,z_{j}[\right)
\end{align*}
where $\exp(\pm./h)$ denotes the function $u\mapsto\exp(\pm u/h)$.
Remark that we used a constant bandwidth $h$, as neither a balloon
bandwidth $h=h_{j}$ nor a sample point bandwidth $h=h_{i}$ can separate
the term $\exp\left(\frac{x_{i}-z_{j}}{h}\right)$ into a product
of a term depending on $i$ only and a term depending on $j$ only.
Note that an intermediate adaptive bandwidth approach of the type
$\frac{x_{i}}{h_{i}}-\frac{z_{j}}{h_{j}}$ would maintain the ability
to separate sources and targets for this kernel.

\subsection{Cosine kernels\label{subsec:Cosine-kernels}}

For the cosine kernel, $K(u)=\frac{\pi}{4}\cos\left(\frac{\pi}{2}u\right)\mathbbm{1}\{\left|u\right|\leq1\}$
and
\begin{align*}
 & \sum_{i=1}^{N}K\!\left(\frac{x_{i}-z_{j}}{h}\right)x_{i}^{p}y_{i}^{q}\mathbbm{1}\left\{ \left|\frac{x_{i}-z_{j}}{h}\right|\leq1\right\} \\
 & =\frac{\pi}{4}\sum_{i=1}^{N}\left\{ \cos\!\left(\frac{\pi}{2}\frac{x_{i}}{h}\right)\cos\!\left(\frac{\pi}{2}\frac{z_{j}}{h}\right)+\sin\!\left(\frac{\pi}{2}\frac{x_{i}}{h}\right)\sin\!\left(\frac{\pi}{2}\frac{z_{j}}{h}\right)\right\} x_{i}^{p}y_{i}^{q}\mathbbm{1}\left\{ z_{j}\!-\!h\leq x_{i}\leq z_{j}\!+\!h\right\} \\
 & =\frac{\pi}{4}\cos\!\left(\frac{\pi}{2}\frac{z_{j}}{h}\right)\mathcal{S}^{p,q}\!\left(\cos\!\left(\frac{\pi}{2}\frac{.}{h}\right),[z_{j}\!-\!h,z_{j}\!+\!h]\right)+\frac{\pi}{4}\sin\!\left(\frac{\pi}{2}\frac{z_{j}}{h}\right)\mathcal{S}^{p,q}\!\left(\sin\!\left(\frac{\pi}{2}\frac{.}{h}\right),[z_{j}\!-\!h,z_{j}\!+\!h]\right)
\end{align*}
where we used that $\cos(\alpha-\beta)=\cos(\alpha)\cos(\beta)+\sin(\alpha)\sin(\beta)$.
In a similar manner, one can define a new kernel based on the hyperbolic
cosine function
\[
K(u)=\frac{1}{4-2\frac{\sinh(\log(2+\sqrt{3}))}{\log(2+\sqrt{3})}}\left\{ 2-\cosh(\log(2+\!\sqrt{3})u)\right\} \mathbbm{1}\{\left|u\right|\!\leq\!1\}
\]
and use the identity $\cosh(\alpha-\beta)=\cosh(\alpha)\cosh(\beta)-\sinh(\alpha)\sinh(\beta)$
to obtain a similar decomposition.

\subsection{Combinations\label{subsec:Combinations}}

Finally, one can combine the polynomial, absolute, and cosine approaches
together to generate additional kernels compatible with fast sum updating.
This combination approach contains the tricube kernel $K(u)=\frac{70}{81}(1-\left|u\right|^{3})^{3}\mathbbm{1}\{\left|u\right|\leq1\}$
(polynomial + absolute value) and the Silverman kernel $K(u)=\frac{1}{2}\exp\left(-\frac{\left|u\right|}{\sqrt{2}}\right)\sin\left(\frac{\left|u\right|}{\sqrt{2}}+\frac{\pi}{4}\right)$
(absolute value + cosine). New kernels can be created by combining
cosine kernels with polynomials, or the three approaches together.
Obtaining the updating equations for these combined kernels is a straight
application of the decomposition tools used in the previous subsections
\ref{subsec:Polynomial-kernels}, \ref{subsec:Absolute-kernels} and
\ref{subsec:Cosine-kernels}.

\section{Stable fast sum updating\label{sec:Stable-fast-sum}}

As observed in \citet{Fan94} and in the numerical section \ref{sec:Numerical},
the numerical rounding errors are invisible when using double-precision
floating-point format. Nevertheless, it is possible to greatly reduce
or remove altogether the residual floating-point rounding errors by
implementing alternative summation algorithms, as discussed in subsection
\ref{subsec:Numerical-stability}. As an illustration, Algorithm \ref{alg:fast1DkregMK}
below shows how to modify the univariate fast sum updating algorithm
\ref{alg:fast1Dkreg} to use the stable M\o ller-Kahan summation
algorithm (\citet{Moller65}, \citet{Linnainmaa74}, \citet{Ozawa83}).
The multivariate case can be adapted in a similar manner. Note that
this stable version multiplies the computational effort by a constant,
and in the multivariate case, the same is true of the memory consumption.

%\clearpage{}
\afterpage{%
\textcolor{white}{\_}
\thispagestyle{empty}
\begin{algorithm2e}[H]
\DontPrintSemicolon
\SetAlgoLined
\vspace{1mm} 
 \KwIn{\\
  X: sorted vector of N inputs $X[1]\leq \ldots \leq X[N]$\\
  Y: vector of N outputs $Y[1], \ldots, Y[N]$\\
  Z: sorted vector of M evaluation points $Z[1]\leq \ldots\leq Z[M]$\\
  H: vector of M bandwidths $H[1], \ldots, H[M]$\\
\vspace{-0.5mm}
\Comment*[l]{Z and H should be such that the vectors Z-H and Z+H are increasing}}  
  \vspace{0.0mm}
  iL = 1 \Comment*[l]{The indices $1\leq iL \leq iR \leq N$ will be such that the current bandwidth}
  iR = 1 \Comment*[l]{$[Z[m]-H[m],Z[m]+H[m]]$ contains $X[iL],X[iL+1],\ldots,X[iR]$}
\vspace{-0.5mm}
  S[$p_1 , p_2$] = 0, $p_1=0,1,\ldots,4$, $p_2=0,1$ \Comment*[l]{Will contain the sum $\sum_{i=iL}^{iR} X[i]^{p_1}\times Y[i]^{p_2}$}
  run[$p_1 , p_2$] = 0, $p_1=0,1,\ldots,4$, $p_2=0,1$ \Comment*[l]{running float-rounding error}
 \For{$m=1,...,M$}{
  \While{(iR$\leq$N) and (X[iR]$<$(Z[m]+H[m]))}{
   \vspace{0.5mm}
   aux[$p_1 , p_2$] = X[iR]$^{p_1} \times$Y[iR]$^{p_2}$ $-$ run[$p_1 , p_2$] , $p_1=0,1,\ldots,4,  p_2=0,1$\\
   \vspace{0.5mm}
   temp[$p_1 , p_2$] = S[$p_1 , p_2$] + aux[$p_1 , p_2$]\\
   \vspace{1.0mm}
   \uIf{abs(S[$p_1 , p_2$])>abs(aux[$p_1 , p_2$])}{
      \vspace{1.0mm}
      run[$p_1 , p_2$] = (temp[$p_1 , p_2$] $-$ S[$p_1 , p_2$]) $-$ aux[$p_1 , p_2$]\\
   }\vspace{-1.0mm}
   \Else{\vspace{-0.5mm}
      run[$p_1 , p_2$] = (temp[$p_1 , p_2$] $-$ aux[$p_1 , p_2$]) $-$ S[$p_1 , p_2$]\\
   \vspace{-0.5mm}}
   S[$p_1 , p_2$] = temp[$p_1 , p_2$]\\
   \vspace{0.5mm}
   iR = iR + 1\\
  }
  \While{(iL$\leq$N) and (X[iL]$<$(Z[m]$-$H[m]))}{
   \vspace{0.5mm}
   aux[$p_1 , p_2$] = $-$X[iL]$^{p_1} \times$Y[iL]$^{p_2}$ $-$ run[$p_1 , p_2$] , $p_1=0,1,\ldots,4,  p_2=0,1$\\
   \vspace{0.5mm}
   temp[$p_1 , p_2$] = S[$p_1 , p_2$] + aux[$p_1 , p_2$]\\
   \vspace{1.0mm}
   \uIf{abs(S[$p_1 , p_2$])>abs(aux[$p_1 , p_2$])}{
      \vspace{1.0mm}
      run[$p_1 , p_2$] = (temp[$p_1 , p_2$] $-$ S[$p_1 , p_2$]) $-$ aux[$p_1 , p_2$]\\
   }\vspace{-1.0mm}
   \Else{\vspace{-0.5mm}
      run[$p_1 , p_2$] = (temp[$p_1 , p_2$] $-$ aux[$p_1 , p_2$]) $-$ S[$p_1 , p_2$]\\
   \vspace{-0.5mm}}
   S[$p_1 , p_2$] = temp[$p_1 , p_2$]\\
   \vspace{0.5mm}
   iL = iL + 1\\
  }\vspace{-0.5mm}
  \Comment*[l]{Here S[$p_1,p_2$]=$\sum_{i=iL}^{iR} X[i]^{p_1} Y[i]^{p_2}$, which can be used to compute}
\vspace{0.5mm}
\Comment*[l]{SK[$p_{1},p_{2}$]=$\sum_{i=iL}^{iR}X[i]^{p_{1}}Y[i]^{p_{2}}\;K\!\left(Z[m],X[i]\right)$}
  \vspace{1mm}
  C0 = 1.0 $-$ Z[m]$^2$/H[m]$^2$; C1 = 2.0 $\times$ Z[m]/H[m]$^2$; C2 = 1/H[m]$^2$\\
  \vspace{0.5mm}
  SK[$p_1 , p_2$] = C0$\times$S[$p_1 , p_2$] + C1$\times$S[$p_1+1 , p_2$] $-$ C2$\times$S[$p_1+2 , p_2$]\\
  \vspace{0.5mm}
  D[m] = 0.75$\times$SK[0,0]/(H[m]$\times$N)\\  
  \vspace{0.5mm}
  R0[m] = SK[0,1]/SK[0,0]\\
\vspace{-0.5mm}
  R1[m]=$\left[\begin{array}{cc} 1 & \mathrm{Z[m]}\end{array}\right]\left[\begin{array}{cc} \mathrm{SK[0,0]} & \mathrm{SK[1,0]}\\ \mathrm{SK[1,0]} & \mathrm{SK[2,0]} \end{array}\right]^{-1}\left[\begin{array}{c} \mathrm{SK[0,1]}\\ \mathrm{SK[1,1]} \end{array}\right]$
  
  \vspace{-2.0mm}
 }
return D, R0, R1\\ 
\vspace{0.5mm}
 \KwOut{\\
D[m]: kernel density estimate of X\\
R0[m]: locally constant regression of Y on X (kernel regression)\\
R1[m]: locally linear regression of Y on X\\
\Comment*[l]{The three estimates D[m], R0[m] and R1[m] are  evaluated at point Z[m] with bandwidth H[m] and Epanechnikov kernel, for each m=1,$\ldots$ ,M}}
\vspace{-0.5mm}
 \caption{Fast univariate kernel smoothing with stable M\o ller-Kahan summation\label{alg:fast1DkregMK}}
 \end{algorithm2e}
}

\section{Multivariate kernel smoothers\label{sec:Multivariate-kernel-smoothers}}

In a multivariate setting, the kernel density estimator \eqref{eq:localdens}
becomes

\begin{equation}
\hat{f}_{\mathrm{KDE}}(z):=\frac{1}{N}\sum_{i=1}^{N}K_{d,H}(x_{i}-z)\label{eq:localdens-d}
\end{equation}

where $x_{i}=\left(x_{1,i},x_{2,i},\ldots,x_{d,i}\right)$, $i\in\{1,2,\ldots,N\}$
are the input points, $z=\left(z_{1},z_{2},\ldots,z_{d}\right)$ is
the evaluation point, and $K_{d,H}(u)=\left|H\right|^{-1/2}K_{d}(H^{-1/2}u)$
is a multivariate kernel with symmetric positive definite matrix bandwidth
$H\in\mathbb{R}^{d\times d}$. Subsection \ref{subsubsec:multivariate-kernel}
discusses the choice of kernel, and Condition \ref{cond:=00005BKernel-support=00005D}
and subsection \ref{subsec:knn-bandwidth} discuss the possibility
of adaptive bandwidth.

The multivariate version of the Nadaraya-Watson kernel regression
estimator \eqref{eq:localreg0} is given by:
\begin{equation}
\hat{f}_{\mathrm{NW}}(z):=\frac{\sum_{i=1}^{N}K_{d,H}(x_{i}-z)y_{i}}{\sum_{i=1}^{N}K_{d,H}(x_{i}-z)}\label{eq:localreg0-d}
\end{equation}

where $y_{i}$, $i\in\{1,2,\ldots,N\}$ are the output points. Finally,
the multivariate version of the locally linear regression \eqref{eq:localreg1}
is given by:
\begin{equation}
\hat{f}_{\mathrm{L}}(z):=\min_{\alpha(z),\beta_{1}(z),\ldots,\beta_{d}(z)}\sum_{i=1}^{N}K_{d,H}(x_{i}-z)\left[y_{i}-\alpha(z)-\sum_{k=1}^{d}\beta_{k}(z)x_{k,i}\right]^{2}\label{eq:localreg1-d}
\end{equation}
By solving the minimization problem \eqref{eq:localreg1-d}, the multivariate
locally linear regression estimate $\hat{f}_{\mathrm{L}}(z)$ is explicitly
given by:
\begin{equation}
\hat{f}_{\mathrm{L}}(z)=\!\left[\!\!\begin{array}{c}
1\\
z_{1}\\
z_{2}\\
\vdots\\
z_{d}
\end{array}\!\!\right]^{\!T}\left[\!\!\begin{array}{cccc}
{\scriptstyle \!\!\stackrel[i=1]{N}{\sum}K_{d,H}(z,x_{i})} & {\scriptstyle \!\!\!\stackrel[i=1]{N}{\sum}x_{1,i}K_{d,H}(z,x_{i})} & \!\!\cdots\!\! & {\scriptstyle \!\!\!\stackrel[i=1]{N}{\sum}x_{d,i}K_{d,H}(z,x_{i})}\\
{\scriptstyle \stackrel[i=1]{N}{\sum}x_{1,i}K_{d,H}(z,x_{i})} & {\scriptstyle \!\stackrel[i=1]{N}{\sum}x_{1,i}x_{1,i}K_{d,H}(z,x_{i})} & \!\!\cdots\!\! & {\scriptstyle \!\stackrel[i=1]{N}{\sum}x_{1,i}x_{d,i}K_{d,H}(z,x_{i})}\\
\!\vdots & \!\vdots & \!\!\ddots\!\! & \!\vdots\\
{\scriptstyle \stackrel[i=1]{N}{\sum}x_{d,i}K_{d,H}(z,x_{i})} & {\scriptstyle \!\stackrel[i=1]{N}{\sum}x_{d,i}x_{1,i}K_{d,H}(z,x_{i})} & \!\!\cdots\!\! & {\scriptstyle \!\stackrel[i=1]{N}{\sum}x_{d,i}x_{d,i}K_{d,H}(z,x_{i})}
\end{array}\!\!\right]^{\!-1}\left[\!\!\begin{array}{c}
{\scriptstyle \!\!\stackrel[i=1]{N}{\sum}y_{i}K_{d,H}(z,x_{i})}\\
{\scriptstyle \stackrel[i=1]{N}{\sum}y_{i}x_{1,i}K_{d,H}(z,x_{i})}\\
\vdots\\
{\scriptstyle \stackrel[i=1]{N}{\sum}y_{i}x_{d,i}K_{d,H}(z,x_{i})}
\end{array}\!\!\right]\label{eq:localreg1-d-explicit}
\end{equation}
To sum up, computing $\hat{f}_{\mathrm{KDE}}(z)$ requires one sum,
computing $\hat{f}_{\mathrm{NW}}(z)$ requires two sums, and finally
one can check that computing $\hat{f}_{\mathrm{L}}(z)$ requires a
total of $(d+1)(d+4)/2$ sums.

Remark that this paper focuses on the three kernel smoothers \eqref{eq:localdens-d},
\eqref{eq:localreg0-d} and \eqref{eq:localreg1-d-explicit}, but
more general kernel smoothers can be implemented with the same fast
multivariate sum updating algorithm described in this paper. For example,
beyond the locally linear regression \eqref{eq:localreg1-d-explicit},
one can consider locally quadratic or locally polynomial regressions.
Another example is to use the matrices in \eqref{eq:localreg1-d-explicit}
to implement more general regressions that ordinary least squares,
for example penalized regressions such as locally linear Ridge regression
or locally linear Lasso regression.

\section{Fast bivariate sweeping algorithm\label{subsec:Fast-bivariate-sweeping}}

\begin{figure}[h]
\begin{centering}
\includegraphics[width=0.75\paperwidth]{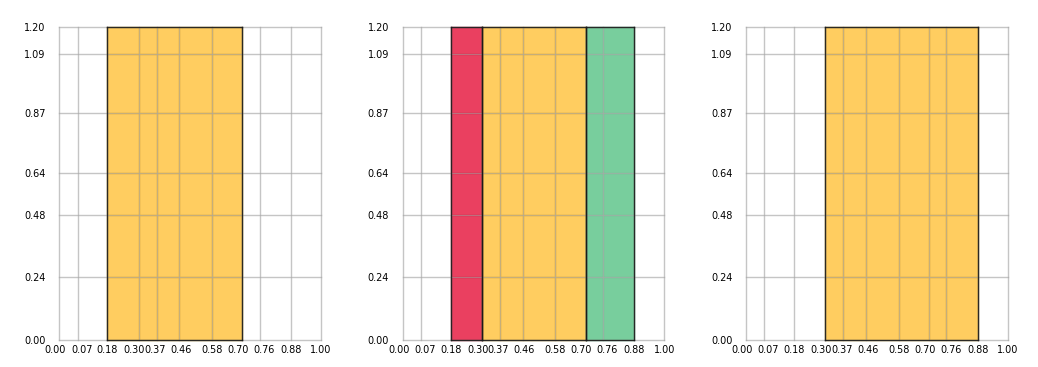}
\par\end{centering}
\caption{Bivariate fast sum updating: outer loop\label{fig:2D_updating_1}}
\end{figure}

\begin{figure}[h]
\begin{centering}
\includegraphics[width=0.75\paperwidth]{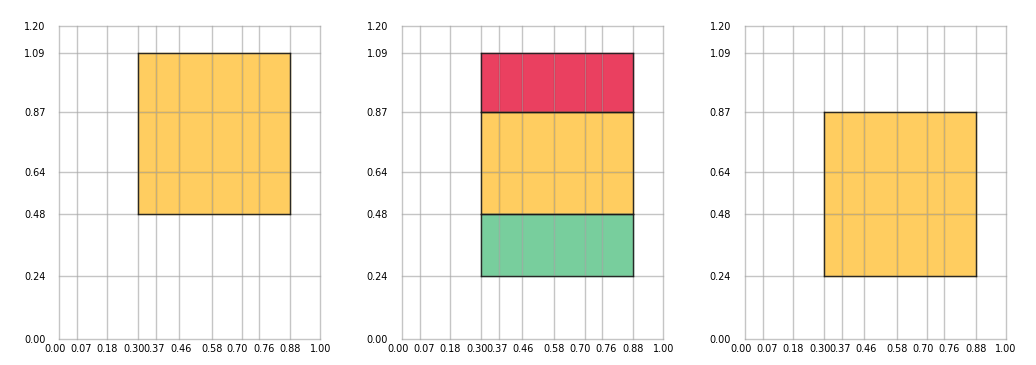}
\par\end{centering}
\caption{Bivariate fast sum updating: inner loop\label{fig:2D_updating_2}}
\end{figure}

This Appendix illustrates the fast bivariate sweeping algorithm \ref{alg:fast2Dkreg}
with the help of Figures \ref{fig:2D_updating_1} and \ref{fig:2D_updating_2}.
As explained in subsection \ref{subsec:Fast-multivariate-sweeping},
we start from $j_{1}=1$ and $\mathcal{T}_{1,l_{2}}^{\mathrm{idx}}=\sum_{l_{1}=L_{1,1}}^{R_{1,1}}\mathcal{S}_{l_{1},l_{2}}^{\mathrm{idx}}$
and iteratively increment $j_{1}$ and update $\mathcal{T}_{1,l_{2}}^{\mathrm{idx}}$
using equation \eqref{eq:2d_updating_1}. Figure \ref{fig:2D_updating_1}
illustrates this fast sum updating in the first dimension. The partition
contains $m_{1}-1=10$ columns and $m_{2}-1=6$ rows. Each rectangle
in the partition is associated with its sum $\mathcal{S}_{l_{1},l_{2}}^{\mathrm{idx}}$.
On the left-side picture, the orange segment on each row $l_{2}$
corresponds to the sum $\sum_{l_{1}=L_{1,j_{1}-1}}^{R_{1,j_{1}-1}}\mathcal{S}_{l_{1},l_{2}}^{\mathrm{idx}}$.
The middle picture represents the fast sum updating \eqref{eq:2d_updating_1}:
for each row $l_{2}$, the green sum $\sum_{l_{1}=R_{1,j_{1}-1}+1}^{R_{1,j_{1}}}\mathcal{S}_{l_{1},l_{2}}^{\mathrm{idx}}$
is added to $\sum_{l_{1}=L_{1,j_{1}-1}}^{R_{1,j_{1}-1}}\mathcal{S}_{l_{1},l_{2}}^{\mathrm{idx}}$
and the red sum $\sum_{l_{1}=L_{1,j_{1}-1}}^{L_{1,j_{1}}-1}\mathcal{S}_{l_{1},l_{2}}^{\mathrm{idx}}$
is subtracted from it. The right-side picture show the result of the
fast sum updating: the orange segment on each row $l_{2}$ corresponds
to the updated sum $\sum_{l_{1}=L_{1,j_{1}}}^{R_{1,j_{1}}}\mathcal{S}_{l_{1},l_{2}}^{\mathrm{idx}}$.

We next turn to the inner loop over $j_{2}$ and $\mathcal{T}_{2}^{\mathrm{idx}}$.
In a similar manner, we start from $j_{2}=1$ and the initial sum
$\mathcal{T}_{2}^{\mathrm{idx}}=\sum_{l_{2}=L_{2,1}}^{R_{2,1}}\mathcal{T}_{1,l_{2}}^{\mathrm{idx}}$,
and iteratively increment $j_{2}$ and update $\mathcal{T}_{2}^{\mathrm{idx}}$
using equation \eqref{eq:2d_updating_2}. Figure \ref{fig:2D_updating_2}
illustrates this fast sum updating in the second dimension. On the
left-side picture, for each row $l_{2}$, the orange segment is associated
with its sum $\mathcal{T}_{1,l_{2}}^{\mathrm{idx}}$. The middle picture
represents the fast sum updating \eqref{eq:2d_updating_2}: the green
sum $\sum_{l_{2}=R_{2,j_{2}-1}+1}^{R_{2,j_{2}}}\mathcal{T}_{1,l_{2}}^{\mathrm{idx}}$
is added to $\sum_{l_{2}=L_{2,j_{2}}}^{R_{2,j_{2}}}\mathcal{T}_{1,l_{2}}^{\mathrm{idx}}$
and the red sum $\sum_{l_{2}=L_{2,j_{2}-1}}^{L_{2,j_{2}}-1}\mathcal{T}_{1,l_{2}}^{\mathrm{idx}}$
is subtracted from it. The right-hand side picture show the result
of this second fast sum updating: the orange hypercube is associated
with the updated sum $\sum_{l_{2}=L_{2,j_{2}}}^{R_{2,j_{2}}}\mathcal{T}_{1,l_{2}}^{\mathrm{idx}}=\sum_{l_{1}=L_{1,j_{1}}}^{R_{1,j_{1}}}\sum_{l_{2}=L_{2,j_{2}}}^{R_{2,j_{2}}}\mathcal{S}_{l_{1},l_{2}}^{\mathrm{idx}}$.
Using Lemma \ref{lem:sum_decomposition} (equation \eqref{eq:sum_partition_slim}),
this sum is equal to $\mathcal{S}_{\mathbf{k}}^{\mathbf{p},q}([z_{j}-h_{j},z_{j}+h_{j}])$
which can be used to compute the kernel sums $\mathbf{S}_{j}=\mathbf{S}_{k_{1},k_{2},j}^{p_{1},p_{2},q}$
using equation \eqref{eq:parabolic-kernel-development-multivariate},
from which the bivariate kernel smoothers (kernel density estimator
\eqref{eq:localdens-d}, kernel regression \eqref{eq:localreg0-d},
locally linear regression \eqref{eq:localreg1-d}) can be computed.

\end{document}